\documentclass[11pt]{article}
\usepackage[utf8]{inputenc}
\usepackage{amsfonts,epsfig}
\usepackage[hyphens]{url}
\usepackage{hyperref}
\usepackage{breakurl}
\usepackage[authoryear]{natbib}

\usepackage{geometry} 
\usepackage{sectsty} 
\usepackage[normalem]{ulem} 
\sectionfont{\sffamily\bfseries\upshape\large}
\subsectionfont{\sffamily\bfseries\upshape\normalsize} 
\subsubsectionfont{\sffamily\bfseries\upshape\normalsize} 
\makeatletter
\usepackage{graphicx}
\usepackage{wrapfig}
 \usepackage{graphicx,fullpage,amsmath,multicol,multirow,amssymb,amsbsy,pifont}
 \usepackage{setspace}
 \usepackage{epsfig}
 \usepackage{amsmath, amsthm, amssymb, bm}
\numberwithin{figure}{section}
\numberwithin{table}{section}
\numberwithin{equation}{section}
 \usepackage{color}
\usepackage{graphics}
\usepackage{wrapfig}
\renewcommand\@seccntformat[1]{\csname the#1\endcsname.\quad}

\makeatletter
\def\@maketitle{%
  \begin{center}%
  \let \footnote \thanks
    {\large \@title \par}%
    {\normalsize
      \begin{tabular}[t]{c}%
        \@author
      \end{tabular}\par}%
    {\small \@date}%
  \end{center}%
}
\makeatother

\title{\bf Bayesian hierarchical weighting adjustment and survey inference\vspace{.1in}}
\author{Yajuan Si\footnote{Survey Research Center, Institute for Social Research, University of Michigan, Ann Arbor, corresponding author; yajuan@umich.edu},  Rob Trangucci\footnote{Department of Statistics, University of Michigan}, Jonah Sol Gabry\footnote{Department of Statistics, Columbia University}, and Andrew Gelman\footnote{Departments of Statistics and Political Science, Columbia University}\vspace{.1in}}

\begin{document}
\maketitle

\begin{abstract}
We combine weighting and Bayesian prediction in a unified approach to survey inference. The general principles of Bayesian analysis imply that models for survey outcomes should be conditional on all variables that affect the probability of inclusion. We incorporate all the variables that are used in the weighting adjustment under the framework of multilevel regression and poststratification, as a byproduct generating model-based weights after smoothing. We improve small area estimation by dealing with different complex issues caused by real-life applications to obtain robust inference at finer levels for subdomains of interest. We investigate deep interactions and introduce structured prior distributions for smoothing and stability of estimates. The computation is done via Stan and is implemented in the open-source R package {\em rstanarm} and available for public use. We evaluate the design-based properties of the Bayesian procedure. Simulation studies illustrate how the model-based prediction and weighting inference can outperform classical weighting. We apply the method to the New York Longitudinal Study of Wellbeing. The new approach generates smoothed weights and increases efficiency for robust finite population inference, especially for subsets of the population.


\end{abstract}
\section{Introduction}

\subsection{Background}
Design-based and model-based approaches have long been contrasted in survey research~\citep{little04-model}. The former automatically takes into account survey design, while the latter can yield robust inference for small sample estimation. \cite{rao:ss11} provides an appraisal of frequentist and Bayesian methods on survey sampling practice. Classical design-based approaches use weights to adjust the sample to the population; see \cite{samsi:review17} for a review of various weighted estimators for a population mean. However, classical survey weighting usually relies on many user-defined choices so that the process of weighting can be difficult to codify in real-world surveys \citep{gelman07}. The Bayesian approach for finite population inference~\citep{ghosh:meeden:97} allows prior information to be incorporated, when appropriate, but is subject to model misspecification.

In the present paper we combine Bayesian prediction and weighting in a unified approach to survey inference, applying scalable and robust Bayesian regression models to account for complex design features under the framework of multilevel regression and poststratification (MRP, \cite{gelman:little:97,park:gelman:bafumi-04,Ghitza:gelman-13,bnfp:ba15}). MRP adjusts for complex design and response mechanisms and improves small area estimation~\citep{fay:herriot79,rao15}. We deal with different complex issues caused by real-life applications and much finer levels for subdomain inference of interest. Our method yields efficient and valid finite population inference, especially for subgroups, and constructs model-based weights after smoothing. 

The contributions of this paper are two folded: 1) as innovative Bayesian methodology developments we develop a new structured prior setting to handle high-order interaction terms; and 2) to improve survey research and operation, we combine Bayesian prediction and weighting as a unified approach to survey inference, accounting for design features in the Bayesian modeling. We generalize MRP for finite population inferences and construct stable and calibrated model-based weights to solve the problems of classical weights. We disseminate the R package {\tt rstanarm} implementing the proposed methods for public use, promoting the model-based approaches in survey research and operational practice. More importantly, the paper builds the groundwork to use MRP in the survey weighting adjustment and data integration, for example, to make inferences with nonprobability surveys. Our proposed methods offer one important and practical tool for designing and weighting survey samples~\citep{valliant-book18}. 

\subsection{Framework}
For a finite population of $N$ units, we denote the variable of interest as $y=(y_1,\dots,y_N)$ and the inclusion indicator variable as $I=(I_1,\dots,I_N)$, where $I_i=1$ if unit $i$ is included in the sample and $I_i=0$ otherwise. Here, inclusion refers to selection and response. The general inference framework considers the joint distribution for $I$ and $y$. Design-based inference considers the distribution of $I$ and treats $y$ as fixed. Under probability sampling, model-based inferences can be based on the distribution of $y$ alone given the variables that affect the inclusion mechanisms are included in the model~\citep{royall68}, that is, under the ignorable inclusion mechanism when the distribution of $I$ given $y$ is independent of the distribution of $y$~\citep{rubin76,rubin83-pi}. 

To account for the factors that affect inclusion, classical design weights adjust for unequal probabilities of sampling, with subsequent weighting accounting for coverage problems and nonresponse during data collection or data cleaning. Classical weights are thus generated as a product of multiple adjustment factors:  inverse probability of selection, inverse propensity score of response, and poststratification (also called calibration or benchmarking; \cite{hs79}).  Each of these adjustments can be approximate when the probability of selection, the probability of response, or population totals are estimated from data.  Beyond any approximation issues, even if the inclusion model is known exactly, extreme values of weights will cause high variability and then inferential problems, especially when the weights are weakly correlated with the survey outcome variable~\citep{Rao66a,Rao66b,hajek71,modelass-sarndal92}. When the weighting process involves poststratification or nonresponse adjustment---where the weights themselves are random variables---the variance estimation will be different from the cases only with fixed design weights. It is nontrivial to analytically derive a variance estimator under the multi-stage weighting adjustment or complex sampling design. 

In practice, the construction of survey weights requires somewhat arbitrary decisions of the selection of variables and interactions, pooling of weighting cells, and weight trimming. It can be unclear whether and how to incorporate auxiliary information~\citep{groves:couper:98JOS}. Discussion of smoothing and trimming in the survey weighting literature \cite[e.g.][]{potter88,trim-potter90,modeltrim-elliottandlittle00,elliott07,elliot:JOS16} has focused on estimating the finite population total or mean, with less attention to subdomain estimates. \cite{beaumont08} proposes to regress weights on the survey variables and use the predicted values as smoothed weights, where the direction is inspiring but tangential to the inference objective where good inference properties are desired for the survey variable of interest rather than the weights. Borrowing information on survey outcomes potentially increases efficiency and calls for a general framework.

\cite{gelman07} recommends regression models including as covariates any variables that affect selection and response, including stratification variables, clusters, and auxiliary information. Any of these approaches can be sensitive to prior specification for stable estimation; this is the model-based counterpart to the decisions required for smoothing or trimming classical survey weights. Flexible prediction techniques, such as spline functions, penalized regression and tree-based models, have been proposed to accommodate model-assisted survey estimation~\citep{modelass-sarndal92,wu:sitter01,model-assist-review-SS17,Toth18}.

Model-based and model-assisted weighting adjustment methods for finite population total estimation have been compared by  \cite{henry:valliant12}. The model-based weighting methods in the superpopulation perspective~\citep{FPSi:Valliant00} use predictions from regression models to derive case weights, where the predictions are based on hierarchical linear regression models with various bias corrections~\citep{robustblup:Chambers:JASA93,robust:Firth:JRSSB98}. Based on the finite population total estimation, model-assisted methods derive case weights mainly from calibration on benchmark variables~\citep{calibration:kott09} via the generalized regression estimator (GREG,~\cite{greg92})]. However, the case weights derived from regression predictions can be highly variable and even negative and may damage some domain estimates. Model-based approaches play a vital role in small area estimation but are subject to misspecification and need new developments when the number of domains is large and the inclusion mechanism is not simply random.

To protect against model misspecification, \cite{little83-pi} recommends modeling differences in the distribution of outcomes across classes defined by differential probabilities of inclusion. \cite{bnfp:ba15} construct poststratification cells based on the unique values of inclusion probabilities and build hierarchical models to smooth cell estimates as advocated by \cite{little91, little93}.

We propose to use Bayesian hierarchical models accounting for survey design to generate weights that can be used in design-based inference. The inference is well calibrated and valid with good frequentist properties~\citep{CalibratedBayes:Little11}. For large samples, the inference will parallel with design-based inference. For small samples, the hierarchical model smoothing will stabilize domain estimation and generate robust weighting adjustment.

We use the intrinsic variables that are used for design weight construction, nonresponse adjustment and calibration, assume they are discretized, and construct poststratification cells based on the cross-tabulation. Weights are derived through the regressing survey outcome on variables used for weighting given the poststratification. The inclusion of the outcome variable into weighting and poststratification can avoid model misspecification and potentially increase efficiency~\citep{fuller09}. Multilevel model estimates shrink the cell estimates towards the prediction from the regression model. The MRP framework combines multilevel regression and poststratification, accounts for design features in the Bayesian paradigm, and is then well equipped to handle complex design features. Our proposal distinguishes from the model-based weights in the literature by using the poststratification cell structure and improves by smoothing, thus avoiding negative weight values.

\cite{bnfp:ba15} incorporate weights into MRP, increasing flexibility and efficiency comparing to the pseudo-likelihood approach~\citep{pfeffermann93}. In the present paper we go further, starting from the variables that are used for weighting and constructing model-based weights as byproducts under MRP. We develop a novel prior specification for the regularization to handle potentially large numbers of poststratification cells. The prior setting allows for variable selection and keeps the hierarchical structure among main effects and high-order interaction terms for categorical variables. That is, if one variable is not predictive, then the high-order interactions involved with this variable are also likely to be not predictive, to facilitate model interpretation. \cite{Toth18} use tree-based methods to automatically select poststrata based on auxiliary variables that are potentially correlated with the survey outcome. Our proposed structured prior plays a similar role with the recursive partitioning algorithm to facilitate poststrata selection but improves efficiency by partial pooling. We use the smoothed weights and estimates that are more stable than the regression tree estimator, and the Bayesian framework propagates all sources of uncertainty where \cite{Toth18} ignore the variance for tree growing and use mean squared error to approximate the variance.

We have implemented the computation in the R package {\tt rstanarm}~\citep{rstanarm}. The fully Bayesian inference is realized via Stan~\citep{stan-software:2013,stan-manual:2013}, which uses Hamiltonian Monte Carlo sampling with adaptive path lengths~\citep{hoffman-gelman:2012}.  Stan promotes robust model-based approaches by reducing the computational burden of building and testing new models. The {\tt rstanarm} package allows for efficient Bayesian hierarchical modeling and weighting inference. The codes are publicly available and reproducible. Our developed computation software provides the accessible platform and has the potential to support the unified framework for survey inference.

Section~\ref{problem} introduces the motivating problem of weighting for an ongoing social science survey. We discuss the method in detail Section~\ref{method}. Section~\ref{simulation} describes the statistical evaluation of model-based prediction and weighting inference, and demonstrate the efficiency gains in comparison with classical weighting. We apply the proposal to the real-life survey in Section~\ref{application}. Section~\ref{discussion} summarizes the improvement and discusses further extension.

\section{Motivating application}
\label{problem}
    
Our methodological research is motivated by operational weighting practice for ongoing surveys. Our immediate goal is to construct weights for the New York City Longitudinal Study of Wellbeing (LSW; \cite{RHweighting,RHreport}), a survey organized by the Columbia University Population Research Center, aiming to provide assessments of income poverty, material hardship, and child and family wellbeing of city residents.

We use the LSW as an example to illustrate practical weighting issues and our proposed improvement, with the understanding that similar concerns arise in other surveys. The survey includes a phone sample based on random digit dialing and an in-person respondent-driven sample of beneficiaries from Robin Hood philanthropic services and their acquaintances. We focus on the phone survey here as an illustration. The LSW phone survey interviews 2,002 NYC adult residents, including 500 cell phone calls and 1502 landline telephone calls, where half of the landline samples are from low-income areas defined by zipcode information. The collected baseline samples are followed up every three months. We match the samples to the 2011 American Community Survey (ACS) records for NYC. The discrepancies are mainly caused by the oversampling of the low-income neighborhoods and nonresponse. 

The baseline weighting process~\citep{RHweighting} adjusts for unequal probability of selection, coverage bias, and nonresponse. Classical weights are products of estimated inverse probability of inclusion and raking ratios \citep{gr-rake93}. However, practitioners have to make arbitrary or subjective choices on the selection and values of weighting factors. For example, to construct weights for individual adults, we have to weight up respondents from large households, as just one adult per sampled household is included in the sample. \cite{gelman:little:98} recommend the square root of the ratio of household sizes to family sizes for this weighting adjustment because using household sizes as weights~\cite[for example,][]{acsweighting2014} tend to overcorrect in telephone surveys. The raking operation procedure in practice adjusts for socio-demographic factors without tailoring for particular surveys.
    
The survey organizers are interested in the aspects of life quality of city residents, such as the percentage of children who live under poverty and material hardship. Thus, it is important to get accurate estimates for subpopulations. We would like to develop an objective procedure and let the collected survey data determine the weighting process. The basic principle is to adjust for all variables that could affect the selection and response into weighting. Ideally, we expect that variables used for weighting should include phone availability (number of landline/cell phones and duration with interrupted service), family structure, household structure, socio-demographics and potentially their high-order interaction terms. However, the ACS records only provide information on family size, age, ethnicity, sex, education and poverty gap (a family poverty measure). Meanwhile, considering the substantive analysis goal, the variables describing the number of elder people in the family, the number of children in the family, and the family size, as well as their interactions with poverty gap are recommended by the survey organizers to be included into the weighting process to balance the distribution discrepancy with the population. 
    
To generate classical weights, we select the raking factors that could affect the selection and response, including sex, age, education, ethnicity, poverty gap, the number of children in the family, the number of elder people in the family, the number of working aged people in the family, the two-way interaction between age and poverty gap, the two-way interaction between the number of persons in the family and poverty gap, the two-way interaction between the number of children in the family and poverty gap, and the two-way interaction between the number of elder people in the family and poverty gap. We collect the marginal distributions from ACS and implement raking adjustment. The generated weights have to be trimmed due to some extreme values.
        
However, it is possible that the subjective weighting adjustment includes some variables or interactions that are not essentially predictive or does not take account for all the important factors that could be of substantive interest later. The raking adjustment assumes that these factors are independent. This will cause biased domain inference bases on the cross-tabulation if the correlation structure in the sample is different from that in the population. Ideally, we should match based on the joint distribution of these weighting related variables. However, small cell sizes or empty under the deep interactions will lead to extremely large weights that need cell collapsing. 
        
The problems we face with classical weighting for the LSW baseline survey are reflective of problems for most operational weighting practice in real-life surveys, which are often complicated with complex designs, longitudinal structure or multi-stage response mechanisms. The ad-hoc decisions that often go into classical weighting schemes can result in different practitioners generating different sets of weights for the same survey. In order to avoid the need for subjectivity, it is important to propose a model-based weighting procedure that allows the data to select weighting factors. We would like to incorporate these variables used for weighting into the model for survey outcomes for efficiency gains, model their high-order interaction terms under regularized prior setting and generate the weights that can be equally treated as classical weights. A large number of variables used for weighting and deep interactions will cause small weighting cells based on the cross-tabulation. The small weighting cells call for statistical adjustment for smoothness and stability. 
        
MRP have achieved success for domain estimation at much finer levels. Borrowing the strength of hierarchical modeling framework with an informative prior distribution, we should be able to obtain the estimate after smoothing the sparse cells. Poststratification via census information will match the estimate from the sample to the population. The combination of regression and poststratification is similar to the endogenous poststratification concept~\citep{Breidt08,Dahlke13}. We introduce the MRP framework in detail.

\section{Method}
\label{method}

\subsection{Multilevel regression and poststratification}
In the basic setting, we are interested in estimating the population distribution of the survey outcome $y$. When the weighting process is transparent, we can directly include the auxiliary variable $X$ into regression modeling for the survey outcome $y$. Here $X$ is a $q$-dimensional vector of variables that affect the sampling design, nonresponse and coverage. Conditional on $X$, the distribution of inclusion indicator $I$ is ignorable. 

The selection of the auxiliary variables and the availability of their joint distributions in the population are the key to success for MRP, and also for all other methods to adjust for the sampling selection and nonresponse bias and yield valid population inferences. We recommend including all variables that potentially affect the sample inclusion, such as design information, paradata, and socio-demographics. One advantage of MRP is to perform variable selection and stabilize weights in contrast with noisy classical weights. 

Another practical challenge is that the population distribution of the calibration variables may be unknown. We obtain the joint population control distribution from ACS in our application study. \cite{wang:gelman14} used the aggregated exit polls, \cite{Zhang15-mrp} used the census tract-level information and Yougov~\citep{Yougov} used the Current Population Survey to directly obtain such information for the poststratification adjustment. In practice, we recommend to obtain the population information either directly from census or large studies with minimal errors or estimated based on available information in related studies. Some auxiliary variables' population distribution may not be available in the census database, such as the number of phones, and we can estimate from other surveys as reference samples.  \cite{reilly:gelman:katz01} applied models to predict the unknown population poststratification information. When marginal distributions are available, \cite{rake:little91} discuss an equivalent model approach for raking and \cite{BayesRake18} develop a Bayes raking estimation in the population cell size estimation. We discuss extensions to develop an integrative framework accounting for the estimation uncertainty of unknown control information in Section~\ref{discussion}. The availability of population control information with high quality and predictive power directly affects inferential validity, either for model-based or design-based approaches.

Under MRP, the auxiliary variables $X$ are discretized, and their cross-tabulation constructs the poststratification cells $j$, with population cell sizes $N_j$ and sample cell sizes $n_j$, where $J$ is the total number of poststratification cells \citep{little91, little93, gelman:little:97,gelmancarlin01}. Then the total population size is $N=\sum_{j=1}^J N_j$, and the sample size is $n=\sum_{j=1}^Jn_j$. 

Poststratification inference is different from design-based inference under stratified sampling by the fact that $n_j$'s are now random functions of the sampling distribution $I$. In the repeated sampling of $I$, there is a nonzero probability that $n_j=0$ for some $j$. The usual resolution of this problem is to condition on $n_j$'s observed in the realized sample, however, the sample inference is not design-unbiased conditionally on $n_j$'s. The MRP framework assumes a model for $n_j$'s to account for the design feature.
    
The poststratification implicitly assumes that the units in each cell are included with equal probability. Suppose $\theta$ is the population estimand of interest, such as the overall or domain means, and it can be expressed as a weighted sum over any subset or domain $D$ of the poststrata,
\begin{align}
    \theta=\frac{\sum_{j\in D}N_j\theta_j}{\sum_{j\in D}N_j},
\end{align}
where $\theta_j$ is the corresponding estimand in cell $j$. 
The proposed poststratified estimator will be of the general form,
\begin{align}
\label{model-based}
    \tilde{\theta}^{\textrm{PS}}=\frac{\sum_{j\in D}N_j\tilde{\theta_j}}{\sum_{j\in D}N_j},
\end{align}
where $\tilde{\theta}_j$ is the corresponding estimate in cell $j$. Various modeling approaches can be used to estimate the cell estimates, such as the flexible nonparametric Bayesian models and machine learning algorithms~\cite{gp-rasmussen06,ml-book09}. Here, we illustrate using a hierarchical regression model.

In practice, survey weights are attached to each unit, even though they are not attributes of individual units. It is natural to generate unit-level weights based on the entire survey design, and use the weighted averages of the form, such as $\tilde{\theta}=\sum_{i=1}^n w_i y_i/\sum_{i=1}^nw_i$. Our goal here is to obtain an equivalent set of unit-level weights $w_i$ through a model-based procedure for the estimation of $\tilde{\theta}^{\textrm{PS}}$ to connect weighting and poststratification. Therefore, regression models can be used to obtain $\tilde{\theta}_j$, poststratification accounts for the population information, and model-based weights are re-derived via the expression \eqref{model-based}.
    
In classical regression models, full poststratification is a special case, where the cell estimates are computed separately for each cell without any pooling effect, i.e., no pooling. For example, if we are interested in the population mean, then the cell means will be used as the cell estimates. Generally, classical regression models are conducted on cell characteristics without going to the extreme fitting separately for each cell. If more interactions among the characteristics are included, the resulted weights become more variable. On the other side, complete pooling ignores the heterogeneity among cells. Hierarchical regression models will smooth the variable estimates under partial pooling. 
    
\cite{gelman07} uses the exchangeable normal model as an illustration and shows that the poststratification estimate $\tilde{\theta}^{\textrm{PS}}$ for population mean can be expressed as a weighted average between the cell means and the global mean, which yields the unit weights, also as a weighted average between the completely smoothed weights, $w_j$ = 1, and the weights from full poststratification, $w_j = (N_j/N)/(n_j /n)$. Hierarchical poststratification is approximately equivalent to shrinkage of weights through the shrinkage of the parameter estimates. The degree of shrinkage goes to zero as the sample size increases, which implies that estimates from the model are close to the truth under the sampling design. However, further developments are necessary to handle a large number of cells and deep interactions, and rigorously evaluate the performance of model-based weights. 

In our application to the LSW study, the variables used for weighting include age (5 categories), ethnicity/race (5 categories), education (4 categories), sex (2 categories), poverty measure (5 categories), family size (4 categories), number of elder people (3 categories) and number of children (4 categories), in the family, and this results in $J=5\times5\times4\times2\times5\times3\times4\times4=48,\!000$ poststrata. The majority of the poststratification cells will be empty or sparse due to the limited sample size (2,002). The sample cell sizes are unbalanced. Often cells are arbitrarily collapsed or combined~\citep{little93} without theoretical justification. Recent model-based weighting smoothing procedures across cells could not handle such sparse cases~\citep{modeltrim-elliottandlittle00}. \cite{elliot:JOS16} introduced a Laplace prior for weight smoothing across a modest number of poststrata based on inclusion probabilities but ignored the variables used for weighting and their hierarchy structure. Using the MRP framework, we account for the variable hierarchy structure to smooth and pool the estimates across the sparse and unbalanced cell sizes with a novel set of prior distributions.

\subsection{Structured prior distribution}

We introduce a structured prior distribution to improve MRP under the sparse and unbalanced cell structures, thus yielding stable model-based survey weights that account for design information. Suppose the population distribution of $X$ is known, that is, we can obtain $N_j$'s from the external data to describe a joint distribution of the variables used for weighting. Extension to unknown $N_j$'s is discussed in Section~\ref{discussion}. In practice, the number of poststratification cells $J$ can be large, even much larger than the sample size $n$. The variables used for weighting could affect the inclusion through a complex relationship or a differential response mechanism. Deep interactions are essential for complex relationship structure, but we cannot include all and have to select the predictive main effects and interactions. 

Suppose the collected survey response is continuous, $y_i$, for $i=1,\dots, n$, and we are interested in the population mean $\bar{Y}$ estimation. We use $(X^{1\top},\dots, X^{J\top})^\top$ to represent the $J\times q$ predictor matrix in the population under the poststratification framework. For illustration, assume a normal distribution,
\begin{align}
\label{normal}
    y_i \sim \textrm{N}(\theta_{j[i]}, \sigma_y^2),
\end{align}
where $j[i]$ denotes the cell $j$ that unit $i$ belongs to. We can also consider unequal variances, allowing the cell scale $\sigma_y$ to vary across cells, indexed as $\sigma_j$. For the prior specification of $\theta_j$, one choice can be  $\theta_j=X^j\beta$, and $\beta$ is assigned with some prior distribution. In the hierarchical regression example of \cite{gelman07}, a multivariate normal distribution is considered, $y_i\sim \textrm{N}(X_i\beta, \Sigma_{y})$ and $\beta \sim \textrm{N}(0,\Sigma_{\beta})$, where the covariates include all main effects and a few selected two-way interactions in $X$ and the covariance matrix $\Sigma_{\beta}$ is diagonal with different scales. However, the model is subject to misspecification, and the generated weights could be negative.
    
Since $X^j$ consists different level indicators of the $q$ discrete auxiliary variables, we can express the population cell mean $\theta_j$ as
\begin{align}
\label{regression}
    \theta_j=\alpha_0 + \sum_{k\in S^{(1)}}\alpha_{j,k}^{(1)}+\sum_{k\in S^{(2)}}\alpha_{j,k}^{(2)}+\dots+\sum_{k\in S^{(q)}}\alpha_{j,k}^{(q)},
\end{align}
where $S^{(l)}$ is the set of all possible $l$-way interaction terms, and $\alpha^{(l)}_{j,k}$ represents the $k$th of the $l$-way interaction terms in the set $S^{(l)}$ for cell $j$. For example, $\alpha^{(1)}_{j,k}$'s with $k\in S^{(1)}$ refer to the main effects, $\alpha^{(2)}_{j,k}$'s with $k\in S^{(2)}$ being the two-way interaction terms, for cell $j$. This decomposition covers all possible interactions among the $q$ variables. When the cell structure is sparse, variable selection is necessary. In practical application, we recommend the initial inclusion of covariates and interactions with substantive importance and scientific interest in Model \eqref{regression} and perform Bayesian variable selection under the proposed structured prior setting.

We induce structured prior distributions to be able to handle deep interactions and account for their hierarchy structure, where the high-order interaction terms will be excluded if one of the corresponding main effects is not selected. Larger main effects often lead to larger effects of the involved interaction terms. Ideally, more shrinkage should be put on the high-order interactions than that on the main effects, and the prior setting should reflect the nested structure. The challenge embodies the problem in Bayesian inference for group-level variance parameters in an ANOVA structure~\citep{anova:gelman:05, gelman06-prior}. \cite{volfovsky:hoff14} introduce a class of hierarchical prior distributions for interaction arrays that can adapt to the potential similarity between adjacent levels, where the covariance matrix for the high-order interactions is assumed as a Kronecker product of the covariance matrices of main effects after adjusting relative magnitudes. Our proposal extends by inducing more structure among the variance parameters, more shrinkage and smoothing effect to handle an extremely large number of cells with unbalanced sizes than the generally balanced setting in~\cite{volfovsky:hoff14}, and improves the computation performance. 

We start with independent prior distributions on the regression parameters $\alpha$:
\[\alpha_{j,k}^{(l)}\sim \textrm{N}(0, (\lambda_k^{(l)}\sigma)^2),\]
where $\lambda_k^{(l)}$ represents the local scale and $\sigma$ is the global error scale, for $k\in S^{(l)}$ and $l=1,\dots, q$. The error scale is the same across the main effects and high-order interactions, while the local scales are different. The shrinkage effect is induced through the specification of local scales. We assume the local scale of high-order interactions is the product of those for the corresponding main effects after adjusting relative magnitudes.
\[\lambda^{(l)}_k=\delta^{(l)}\prod_{l_0\in M^{(k)}}\lambda^{(1)}_{l_0},\]
where $\delta^{(l)}$ is the relative magnitude adjustment and $M^{(k)}$ is the collection of corresponding main effects that construct the $k$th $l$-way interaction in the set $S^{(l)}$. For example, the local scale of the three-way interaction among age, sex, and education, middle-aged men with college education, will be the product of those for the main effects on age, sex, and education, that is, the product of the local scale parameters for middle-aged, men, and college educated, respectively. 

We use the following hyperpriors on the scale parameters:
\begin{align}
\label{main}
    \nonumber  \mbox{error scale: }&\sigma \sim  \textrm{Cauchy}_{+}(0,1) \\ 
    \nonumber \mbox{local scale for main effects: } & \lambda_k^{(1)} \sim  \textrm{N}_{+}(0,1)\\
     \mbox{local scale for high-order interactions: } & \lambda^{(l)}_k=\delta^{(l)}\prod_{l_0\in M^{(k)}}\lambda^{(1)}_{l_0} \\
    \nonumber  \mbox{relative magnitude for high-order interactions: }&\delta^{(l)}  \sim   \textrm{N}_{+}(0,1) \mbox{, for } l=2,\dots, q.
\end{align}
Here $\textrm{Cauchy}_{+}$ and $\textrm{N}_{+}$ denotes the positive part of the Cauchy and normal distributions, respectively.
\cite{gelman06-prior} proposes the half-Cauchy prior for the scale parameter in hierarchical models, which has the appealing property that it allows scale values arbitrarily close to 0, with heavy tails allowing large values when supported by the data. When $\lambda_k^{(l)}$ is close to 0, the posterior samples of $\alpha_{j,k}^{(l)}$ are shrunk towards 0. The scale parameter for the high-order interaction terms will be 0 if any of the related scale parameters for the main effects is 0. The overall regularization effect is determined by the error scale and the multiplicative scale parameters of the corresponding main effects. We assign a noninformative prior distribution to the intercept term and weakly informative prior distributions to the two global error scale parameters $(\sigma_y, \sigma)$, where $\sigma_y\sim \textrm{Cauchy}_{+}(0,5)$.

The global-local shrinkage prior can stabilize random effects modeling in small area estimation~\citep{priorSAE-Tang18}. Our proposed prior specification features the global-local shrinkage and group selection of all possible level indicators for the same variable, similar to the group lasso~\citep{grouplasso06}. We achieve the goal of variable selection under the similar specification with the Horseshoe prior distribution~\citep{horseshoe10} and improve by setting up group selection and multiplicative scales for high-order interactions for sparsity gains. We introduce weakly informative half-Cauchy prior distributions to error scales and informative half-normal prior distributions to the local scale parameters to improve parameter shrinkage estimation and computation efficiency. When the posterior estimation of the scale parameter is close to 0, indicating the variable is not predictive; post-processing can be done to exclude the variable from poststratification cell construction for dimension reduction. This class of priors allows for variable selection in high dimension and keeps the hierarchical structure among main effects and interactions. 

\cite{hyperprior:Aki16} recommend the prior choice for the global shrinkage hyperparameters based on prior beliefs about the number of nonzero coefficients in the model. The hierarchy setting with correlated variables requires further investigation of such recommendation. We use the default choice $\textrm{Cauchy}_{+}(0,1)$ and
conduct an extensive sensitivity analysis of the hyperparameter specification, where the results do not change.
 
\subsection{Model-based weights}

We can re-express~\eqref{regression} and~\eqref{main} as the exchangeable normal model:    
\begin{align}
\label{theta-summary}
    \theta_j \sim N(\alpha_0, \sigma^2_{\theta})\mbox{, } \ \ \sigma^2_{\theta}=\sum_{l=1}^q\sum_{k\in S^{(l)}}(\lambda_k^{(l)}\sigma)^2.
\end{align}
Conditional on the variance parameters, the posterior mean in the normal model with normal prior distribution is a linear function of data; thus we can determine {\em equivalent weights} $w^*_i$'s so that one can re-express the smoothed estimate $\sum_{j=1}^JN_j/N\tilde{\theta}_j$ as a classical weighted average, $\sum_{i=1}^n w^*_iy_i/\sum_{i=1}^n w^*_i$. Combining the posterior mean estimates for $\theta_j$ and the model-based estimate given in Model~\eqref{model-based}, \cite{gelman07} derives the equivalent unit weights in cell $j$ that can be used classically. 
\begin{align}
\label{model-w}
     w_j\approx \frac{n_j/\sigma^2_y}{n_j/\sigma^2_y+1/\sigma_{\theta}^2}\cdot\frac{N_j/N}{n_j/n} +  \frac{1/\sigma^2_{\theta}}{n_j/    \sigma^2_y+1/\sigma_{\theta}^2} \cdot 1,
\end{align}
where the model-based weight is a weighted average between full poststratification without pooling (weights of $(N_j/N)/(n_j/n)$) and complete pooling (weights equal to 1). The pooling or shrinkage factor is $1/(1 + n_j\sigma^2_{\theta}/    \sigma^2_y)$, which depends on the group and individual variances as well as sample size in the cell. The model-based weights are random variables, and fully Bayesian inference will propagate the corresponding variability. We collect the posterior mean values and treat as the weights that can be used the same as classical weights.

\subsection{Computation}

The Bayesian hierarchical prediction and weighting inference procedure is reproducible and scalable. We implement the proposed structured prior distributions in the open source R package {\tt rstanarm}~\citep{rstanarm}. The computation codes are available online~\citep{code:si} for public use. We present the example code for the real data application in Appendix~\ref{code} to demonstrate the user-friendly and efficient computation interface, where survey practitioners can straightforwardly use and adapt. The fully Bayesian inference is realized via Stan. As open source and user-friendly software, Stan contributes to the wide application of Bayesian modeling. Survey practitioners resist model-based approaches mainly due to computation burden. However, model-based methods are ready to face the new challenges on big survey data, such as unbalanced cell structure, combining multiple surveys and analyzing streaming data. The development of Stan can improve the generalization of the model-based approach and provide the computational platform for the unified survey inference framework.

In our implementation, the Markov chain Monte Carlo samples mix well and the chains converge quickly. The fast computation speed widens the usability of model-based survey inference approaches. The proposed prior specification improves the stability for smoothed weights under partial pooling. We compare the model-based weights with classical weights in Section~\ref{simulation} and~\ref{application} to demonstrate the calibration for design-based properties~\citep{CalibratedBayes:Little11}. Furthermore, we illustrate the proposed improvement for domain estimation under unbalanced and sparse sample cell structure.

\section{Simulation studies}
\label{simulation}

We evaluate the Bayesian procedure by the design-based properties and demonstrate the validity. We consider two main simulation scenarios: a slightly unbalanced structure with a moderate number of poststratification cells and very unbalanced structure with a large number of poststratification cells. We evaluate the statistical validity of the model-based and weighted estimation for the finite population and domain inference to demonstrate the improved capability to solve the classical weighting problems. To illustrate the capability of variable selection and hierarchy maintenance and the resulting efficiency gains, we compare the posterior estimation with that under independent prior setting but without the multiplicative scale constraint, which is similar with Horseshoe prior under group specification, called as independent prior distributions in the paper: $\lambda^{(l)}_k \sim \textrm{N}(0,(\sigma^{(l)}_k)^2)$.

We consider model-based predictions under the structured prior (Str-P) and the independent prior (Ind-P) distributions. For weighted inference, we evaluate the estimation after applying the model-based weights under structured prior (Str-W) setting, model-based weights under independent prior (Ind-W) distributions, weights obtained via raking adjustment (Rake-W), classical poststratification weights (PS-W), and inverse probability of selection weighting (IP-W). We present the graphical diagnosis tools to compare the weights and weighted inference.

We borrow 2011 ACS survey of NYC adult residents treating it as the ``population", and randomly draw samples out of it according to a pre-specified selection model without nonresponse. We collect covariates from ACS and simulate the outcome variable to obtain the true distribution as a benchmark. The details of model specifications for the following scenarios are presented in Appendix~\ref{appendix}. We implement the raking procedure by balancing the marginal distributions of the calibration variables in the selection model and generate the raking weights. The classical poststratification weights $N_j/n_j$'s are obtained by matching the selected sample cell indices with those of the population cells. The selection model can provide the inverse probability of selection weights by matching the sampled unit indices. We also generate model-based weights under independent prior distributions for the main effects and high-order interaction terms of the ACS variables. The generated weights are normalized to average 1 for comparison convenience. 

\subsection{Slightly unbalanced structure}
\label{3var}

We first handle slightly unbalanced structure when the number of poststratification cells and the sample cell sizes are moderate. We implement repeated sampling process to investigate the frequentist properties of model-based predictions and weighted inferences. With little shrinkage effect on high-order interactions, the model-based prediction and weighting with structured prior distributions have similar performance with that under independent prior distributions, while outperforming the classical weighting approaches. 

Assume three variables are included in the selection and outcome models: age, ethnicity, and education. We discretize the three variables in ACS as {\em age} (18--34, 35--44, 45--54, 55--64, 65+), {\em eth} (non-Hispanic white, non-Hispanic black, Asian, Hispanic, other), and {\em edu} (less than high school, high school, some college, bachelor degree or above). The number of poststratification cells is $5\times 5 \times 4 = 100$. We assume the outcome depends on deep interactions, including all the main effects, two-way and three-way interaction terms among the three variables; and the selection indicator depends on the three main effects. The specific values of the coefficients are given in Tables~\ref{s1-response-coef}--\ref{s1-selection-coef} in Appendix~\ref{appendix}. The values are set to reflect the strong correlations between the covariate and dependent variables. And the effects are not necessarily similar across the adjacent factor levels, different from the scenarios in~\cite{volfovsky:hoff14}. The error scale in the outcome model is set as 1, where the true value is always fully recovered from the posterior estimation. The data generation model is different from the estimation model, but the latter is flexible enough to cover the former since the dependency structure will be recovered by the estimation. The proposal is robust against model misspecification.

We repeat the sampling 500 times. The sample sizes vary between 2141 and 2393 with median 2288. Empty sample cells occur with spread-out selection probabilities (ranging from 0.001 to 0.269) over the repeated sampling process. The number of occupied cells in the sample is between 80 and 93 with median 87. The slightly unbalanced cell structure is common in practical surveys with simple and clean sampling design. The population quantities of interest include the overall mean, domain means across the $13\, (=5+4+4)$ marginal levels of three variables and domain mean for nonwhite youths (an example of interaction between age and ethnicity). We examine the absolute value of estimation bias, root mean squared error (RMSE), standard error (SE) approximated by the average value of standard deviations (Ave.\ SD) and nominal coverage rate of the 95\% confidence intervals.

\begin{figure}
\centering
\begin{tabular}{cc}
\includegraphics[width=.475\textwidth]{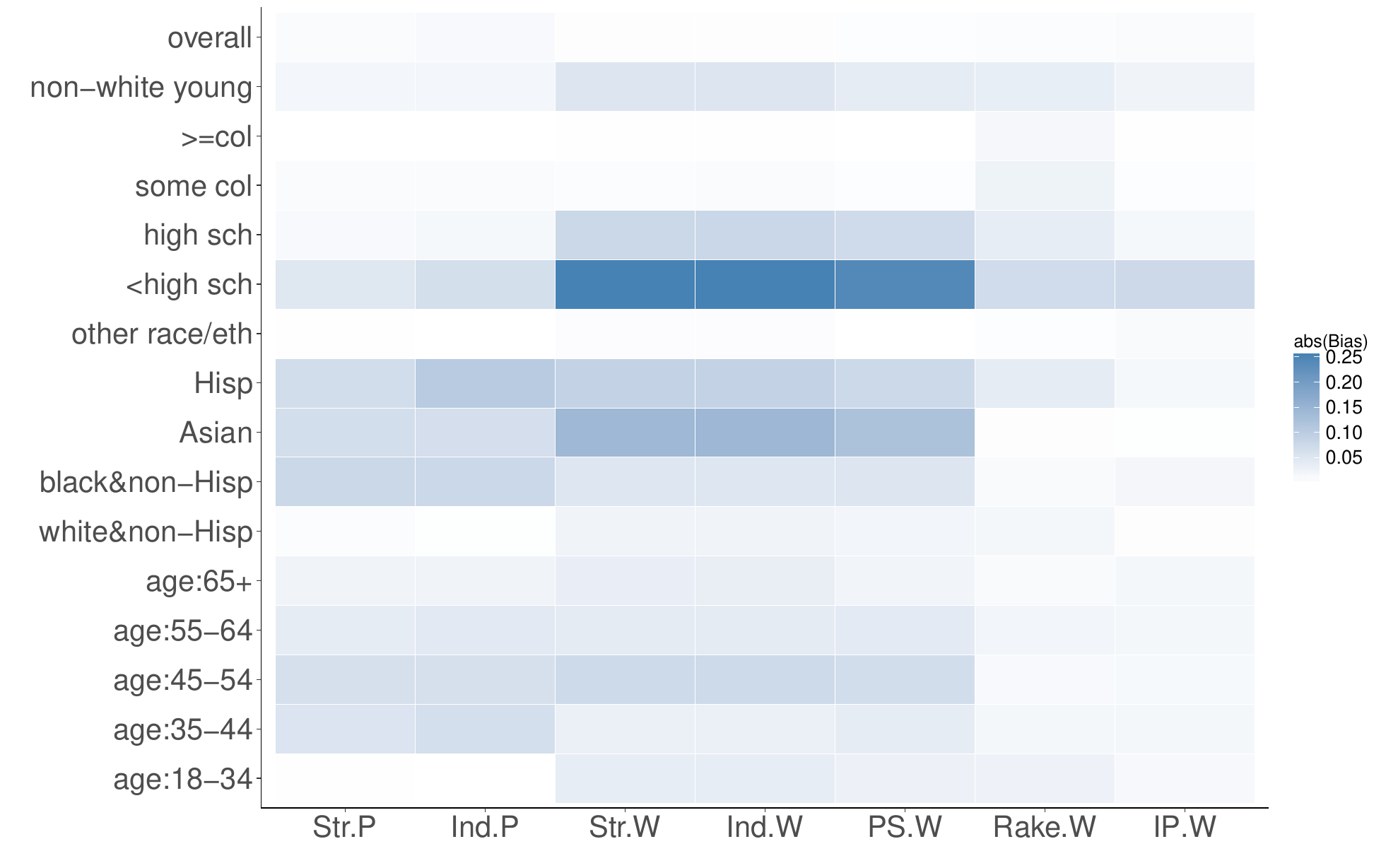}&
\includegraphics[width=.475\textwidth]{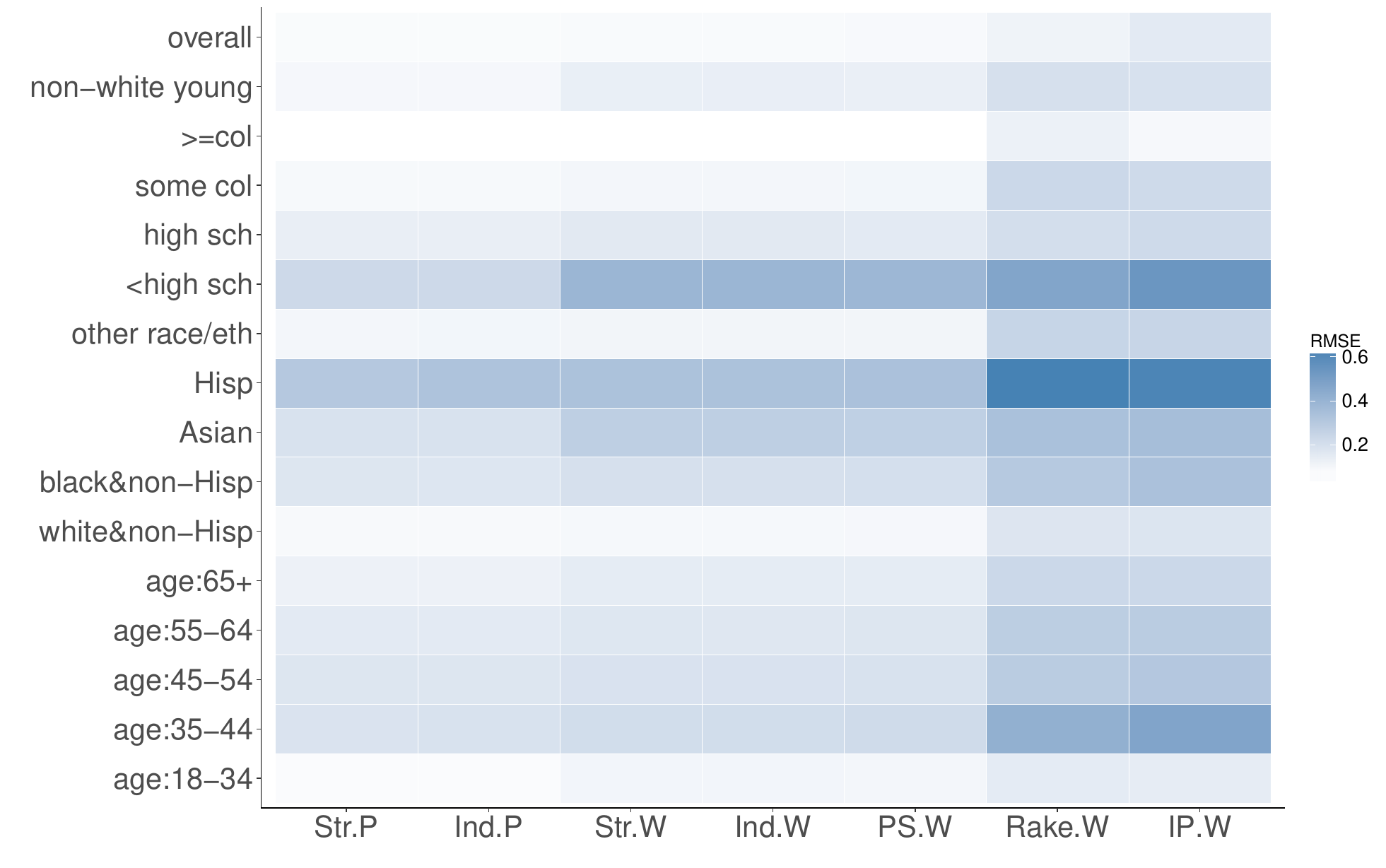}\\
\includegraphics[width=.475\textwidth]{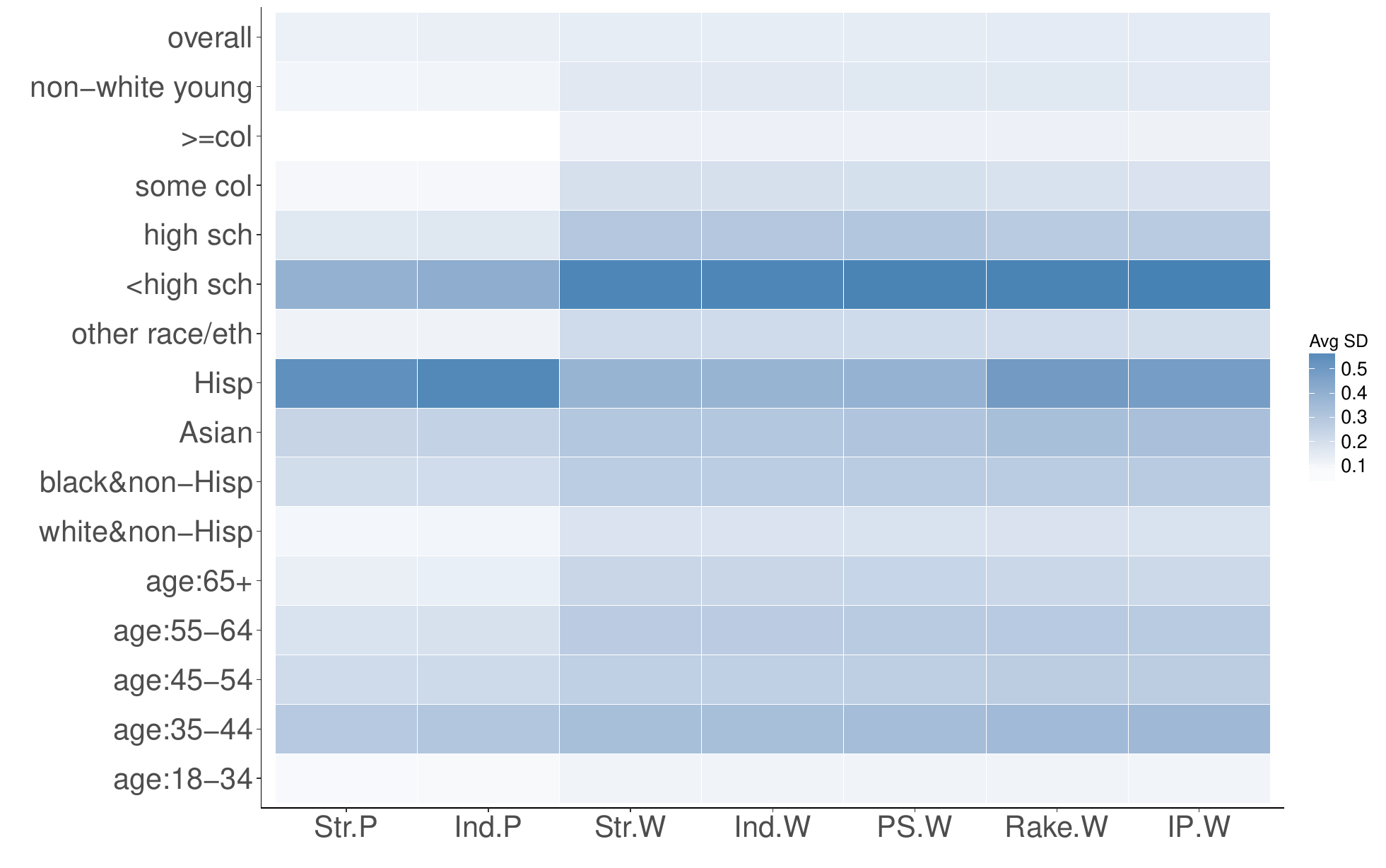}&
\includegraphics[width=.475\textwidth]{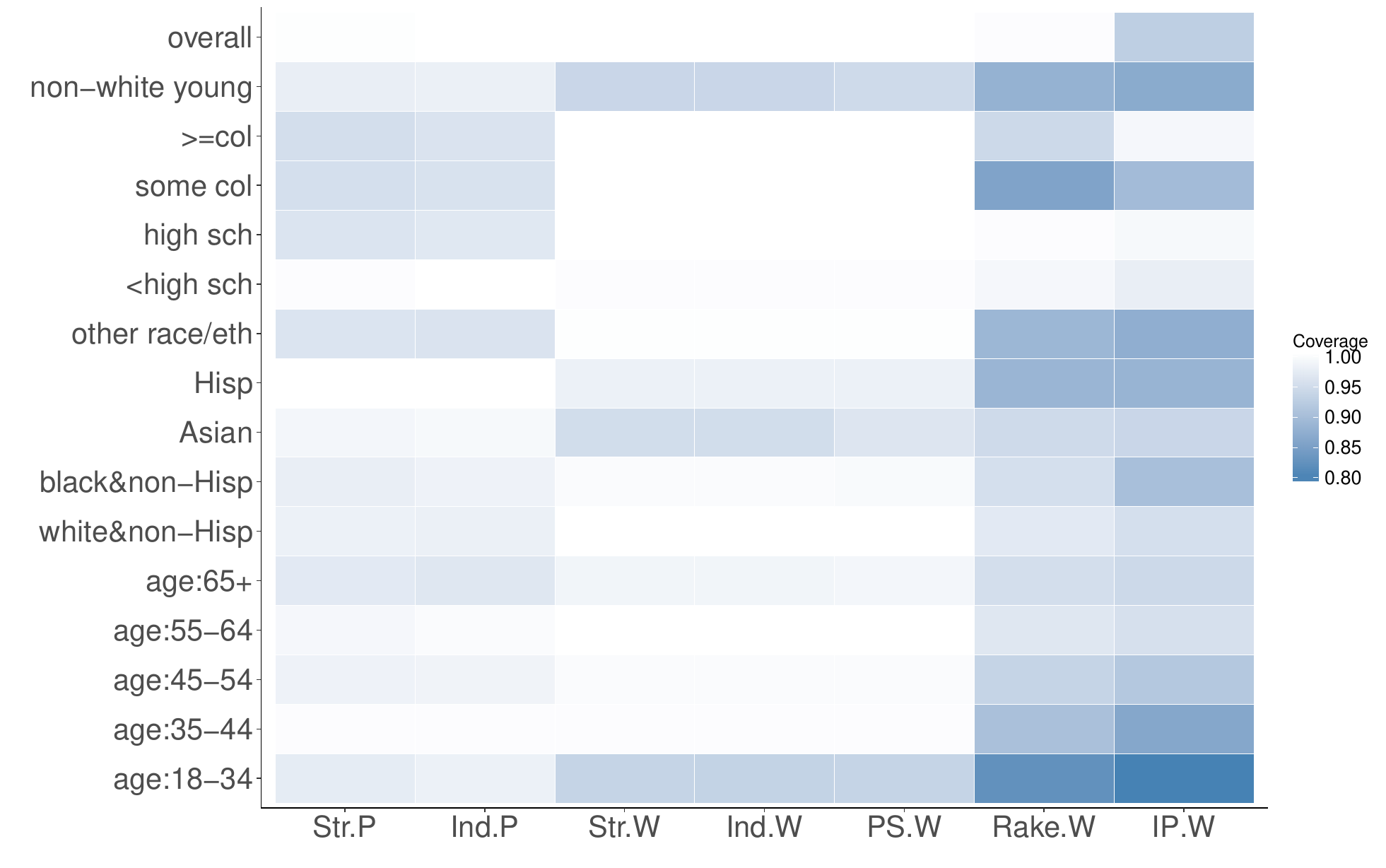}\\
\end{tabular}
\caption{\em Comparison of prediction and weighting performances on the validity of finite population inference under slightly unbalanced design. The y-axis denotes different groups for the mean estimation. The x-axis includes two model-based prediction methods (Str-P, Ind-P), two model-based weighting methods (Str-W, Ind-W), and three classical weighting methods (PS-W, Rake-W, IP-W). Str-P: model-based prediction under the structured prior; Ind-P: model-based prediction under the independent prior distribution; Str-W: model-based weighting under structured prior; Ind-W: model-based weighting under independent prior distribution; Rake-W: weighting via raking adjustment; PS-W: poststratification weighting; and IP-W: the inverse probability of selection weighting. The plots show that the model-based predictions outperform weighting with the smallest RMSE, the smallest SE, reasonable coverage rates, and comparable bias among all the methods. Model-based weighting inference has smaller RMSE and SE but more reasonable coverage rates than that with classical weighting.}
\label{sim1}
\end{figure}

The outputs in Figure~\ref{sim1} show that the model predictions have the smallest RMSE, the smallest SE with reasonable coverage rates, and comparable bias among all the methods. All variables affecting the outcome and selection mechanism are included in the modeling to satisfy the Bayesian principle for ignorable sampling mechanism. The model will predict all the cell estimates including the empty cells in the sample, fully using the population information and poststratification cell structure. The weighting inference is conditional on the observed units within occupied cells, and thus less efficient than the model predictions. Generally, the model-based weighting inference has smaller RMSE and SE but more reasonable coverage rates than that with classical weighting. Raking adjustment is not valid for the domain estimation with large bias, large RMSE, and poor coverage, even though the selection mechanism depends on only the main effects. The inverse probability of selection weighting inference tends to have large SE but low coverage rates, especially for domain estimation. The poststratification weighting inference is close to the model-based weighting estimation since the domain sizes are modestly large. The cell shrinkage effect towards no weighting is small (between 0 and 0.19 with mean 0.05) under slightly unbalanced design. The number of cases who are less than high school educated is small (around 80), resulting in large estimation bias and SE for the weighting inferences, but not in model-based predictions. The model-based predictions stabilize the small area estimation by smoothing, as shown in Table~\ref{sim1_sub_int} that displays the numerical comparison for the subdomain inference. 
\begin{table}[t]
\centering
\label{sim1_sub_int}
\begin{tabular}{rrrrrrrr}
& Str.P & Ind.P & Str.W & Ind.W & PS.W & Rake.W & IP.W \\ 
  \hline
Abs.Bias & 0.02 & 0.02 & 0.05 & 0.05 & 0.04 & 0.03 & 0.02 \\ 
RMSE & 0.07 & 0.07 & 0.11 & 0.11 & 0.10 & 0.17 & 0.17 \\ 
Ave.SD & 0.08 & 0.08 & 0.13 & 0.13 & 0.13 & 0.13 & 0.13 \\ 
Coverage & 0.97 & 0.98 & 0.94 & 0.94 & 0.94 & 0.88 & 0.86 \\ 
\end{tabular}
\caption{Comparison of prediction and weighting performances on the subgroup mean for non-white youth under slightly unbalanced design.}
\end{table}
     
Model prediction performs well and similarly under the structured prior distribution or independent prior distribution. This is expected due to the small shrinkage effect. The cell structure is slightly unbalanced, and the outcome and selection models depend on all the main effects and high-order interaction terms. But the structured prior setting yields more efficient inference than the independent prior setting with smaller SE. This improvement is obvious in the very unbalanced design as shown in the following simulation of Section~\ref{8var}.

Additionally, we considered nine cases with different survey outcome models and sample selection models depending on various predictors as in Table~\ref{s1-design} in Appendix~\ref{appendix}. The specific values of the coefficients are given in Tables~\ref{s1-response-coef}--\ref{s1-selection-coef}. The conclusions are consistent that the model-based prediction and weighting yield more efficient and precise inference than that under classical weighting, in particular for domain estimation.

\subsection{Very unbalanced structure}
\label{8var}

Complex sampling design and response mechanisms tend to create very unbalanced data structures where most poststratification cells are sparse and empty. The proposed structured prior setting brings in strong regularization effect to stabilize the model prediction and improves the estimation efficiency, especially for domain estimation, outperforming the independent prior distributions. The posterior inference on scale parameters can inform variable selection to improve model interpretation. When the main effects are not predictive, neither are the related high-order interactions. However, the posterior inference with independent prior distributions distorts the hierarchical structure between main effects and high-order interactions and hardly informs variable selection. The classical weighting inferences are highly variable in the sparse scenario. 

Following the LSW, we collect eight variables in the 2011 ACS-NYC data that affect sample inclusion: {\em age} (18--34, 35--44, 45--54, 55--64, 65+), {\em eth} (non-Hispanic white, non-Hispanic black, Asian, Hispanic, other), {\em edu} (less than high school, high school, some college, bachelor degree or above), {\em sex} (male, female), {\em pov} (one household income or poverty measure, poverty gap under 50\%, 50--100\%, 100--200\%, 200--300\%, more than 300\%), {\em cld} (0, 1, 2, 3+ young children in the family), {\em eld} (0, 1, 2+ elders in the family), and {\em fam} (1, 2, 3, 4+ individuals in the family). The number of unique cells occupied by this classification is 8874, while the number of poststratification cells constructed by the full cross-tabulation is 48,000. 

In the simulation described in Table~\ref{s2-design} and Table~\ref{s2-design-case2}, the selection probability depends on the main effects of all variables, while the outcome depends on the main effects of five variables. The cell selection probabilities will be clustered, where some cells have the same selection probabilities. The error scale in the outcome model is set as 1. The selection probabilities fall between 0 and 0.90 with average 0.12, and we select 6374 units. Even though the sample sizes are large, the simulation creates a very unbalanced structure. The majority of the cells are empty, and 1096 of 1925 selected cells have one unit. Starting from an estimation model with sparsity, we assume the Model (\ref{regression}) for the cell estimations includes the main effects of the eight variables, eight two-way interactions, and two three-way interactions. These terms are potentially important factors for weighting from the survey organizer's view. Our proposal can provide the insight of variable selection and then facilitate dimension reduction. 

When only the main effects are predictive, the posterior median values under the structured prior setting for the scales of the {\em cld}, {\em eld}, and {\em fam} are small (0.002, 0.003, 0.000), and the posterior median values for the scales of all high-order interactions are close to 0 (with magnitude smaller than or around 0.0001). The posterior mean of the error scale is 0.99 with SE 0.008, close to the true value 1. This is consistent with the simulation design. With independent prior distributions, however, the hierarchical structure between the main effects and high-order interaction terms is ignored. The posterior samples of scale parameters of the high-order interactions can be larger than that of the main effects. It is unclear about their predictive power and then hard to decide which terms to be selected. The posterior samples of the variance parameters under the independent prior distributions tend to be highly variable with heavy tails. For example, the variances of the main effects of age and sex have extremely large sampled values (14496 and 390000) and skewed distributions. For variables with a small number of levels, such as sex, the group-level variance estimation is sensitive to the prior distribution, and the independent prior distribution cannot regularize well. The structured prior distribution performs better by assuming the prior distributions share some common parameter and using more information for estimation and then is able to stabilize the variance estimation. The structured prior setting yields more stable inference than the independent prior, and moreover can facilitate variable selection.  

The proposed structured prior setting suggests that we exclude the nonpredictive main effects and high-order interactions from the regression model for cell estimates, by either post-processing the posterior samples of the corresponding scales and coefficients to be 0 or refitting the updated model. In the simulation design, three variables affect the selection probability but are not related to the outcome. The inclusion of these variables into the regression model will increase the inference variability. The poststratification cell structure accounts for the eight variables to meet the ignorable sampling assumption. A further modification could be the exclusion of the three variables from the poststratification, which could make the assumption of ignorable sampling vulnerable but have efficiency gains. This is a tradeoff between efficiency and robustness that needs balance based on substantive interest. The selection of survey outcome variables in the weighting process needs further investigation, which we will elaborate in Session~\ref{discussion}. We compared the inference with that after excluding the nonpredictive terms and obtained similar outputs for the finite population and domain estimation since the parameter estimates are close to 0 for the nonpredictive terms. Here we present the outputs keeping such variables in the poststratification cell construction and the regression model.

\begin{figure}
\centering
\begin{tabular}{cc}
\includegraphics[width=.475\textwidth]{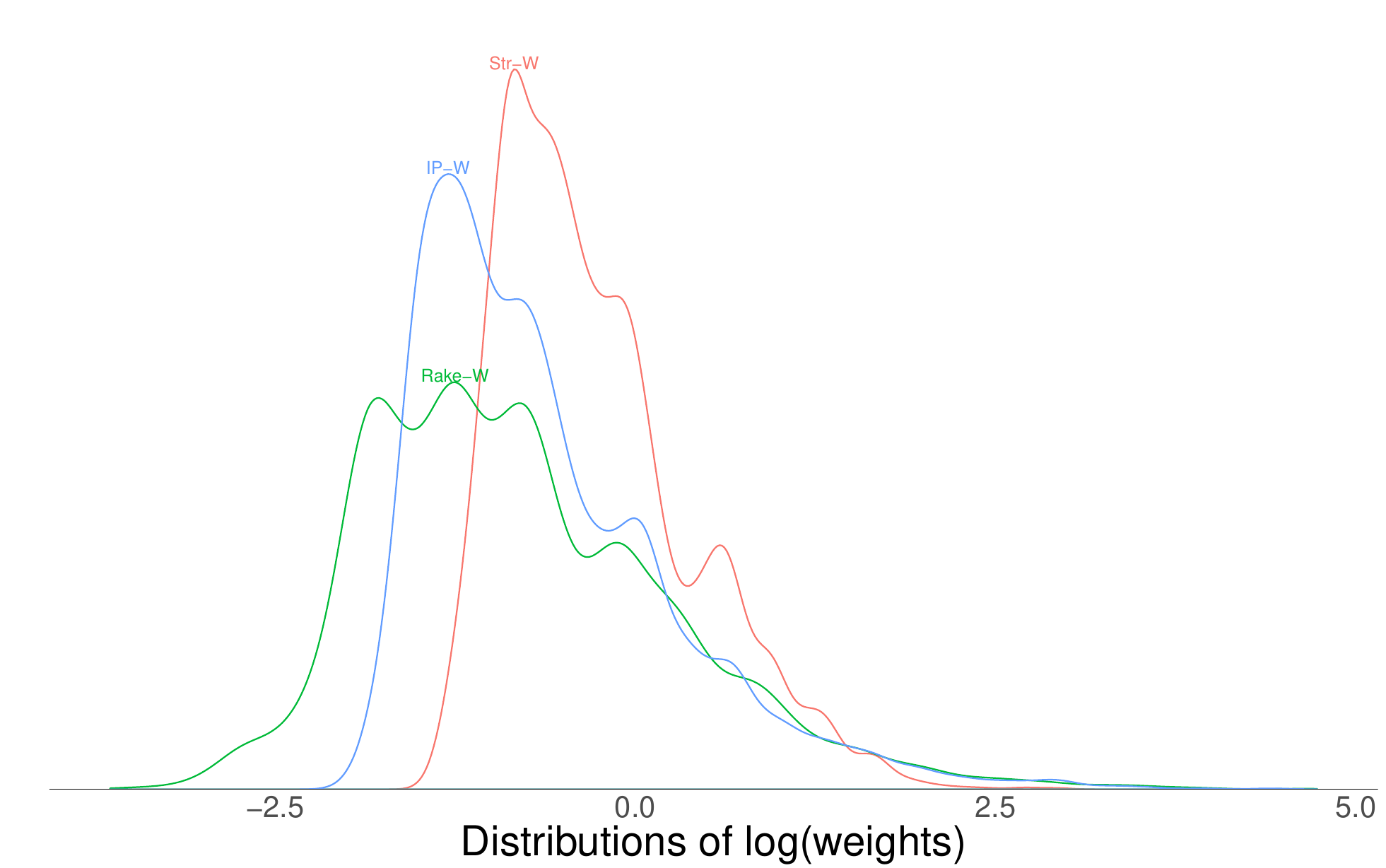}&\includegraphics[width=.475\textwidth]{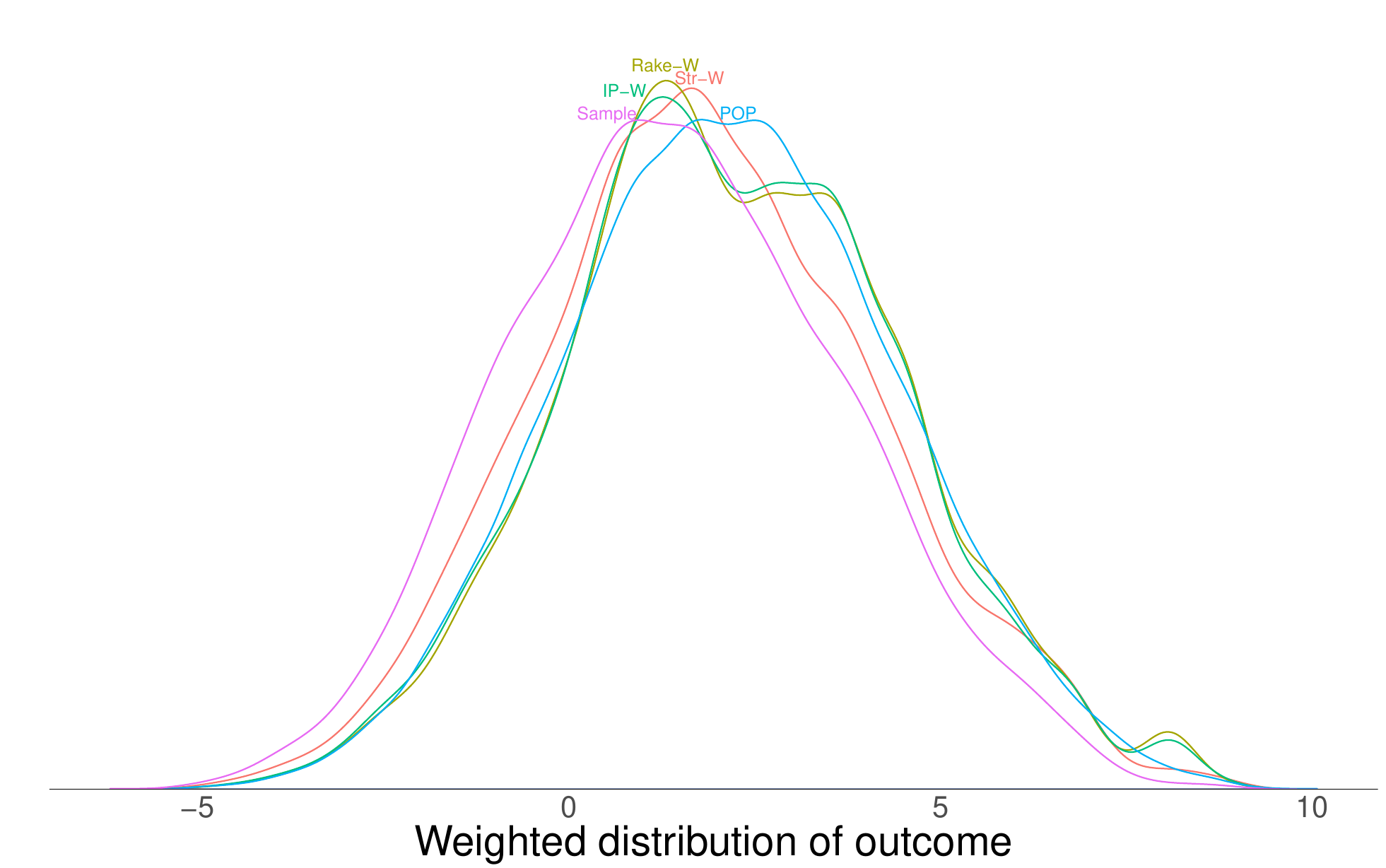}
\end{tabular}
\caption{\em Comparison of generated weights after logarithmic transformation and weighted outcome distributions under a very unbalanced design. Str-W: model-based weighting under structured prior; Rake-W: weighting via raking adjustment; and IP-W: inverse probability of selection weighting. Sample: sample distribution of the outcome; and POP: population distribution of the outcome. The model-based weights are more stable and generate a more smoothed outcome distribution after weighting than the raking weights and the inverse probability of selection weights.}
\label{sim2-weight}
\end{figure}

First, we compare the generated weights by the model-based and classical methods. We collect the posterior samples of generated weights and present the posterior mean as the model-based weights. The model-based weights have smaller variability and narrower range than the classical weights, as shown in Figure~\ref{sim2-weight}. The iterative proportional fitting procedure does not converge after the default 10 iterations that need increasing. We examine the distribution of the outcome after accounting for the weights and compare with the population and sample distribution in the right plot of Figure~\ref{sim2-weight}. The sample distribution differs from the population distribution by underestimating the outcome values. The weighted distribution shifts towards the true population. The outcome distributions after weighting are similar among the model-based and classical methods, and the model-weights generate a smooth distribution of outcomes. This is reasonable as we expect the model-based weights perform similarly with classical weights on point estimation but improve efficiency by reducing the variability. The shrinkage effect under the structured prior distribution is large, between 0.86 and 1.00 with mean 0.90. The very unbalanced cell structure needs a strong smoothing effect across cells. The model-based weights under the structured prior and independent distributions have similar distributions with the poststratification weights, so the latter two sets of weights are omitted in Figure~\ref{sim2-weight}.

We examine the inference for the overall mean and domain means across the marginal levels and for nonwhite young adults. The conclusions are the same as that in Section~\ref{3var}. Model-based prediction outperforms weighting inference with smallest bias and SE. The benefit can be explained by that the model uses the population information for empty cell prediction under regularization. Model-based weighting inference has smaller SE than that with classical weighting. Even when the selection probabilities depend on only main effects, raking yields small bias but performs badly with large SE. 

\begin{figure}
\centering
\begin{tabular}{cc}
\includegraphics[width=.475\textwidth]{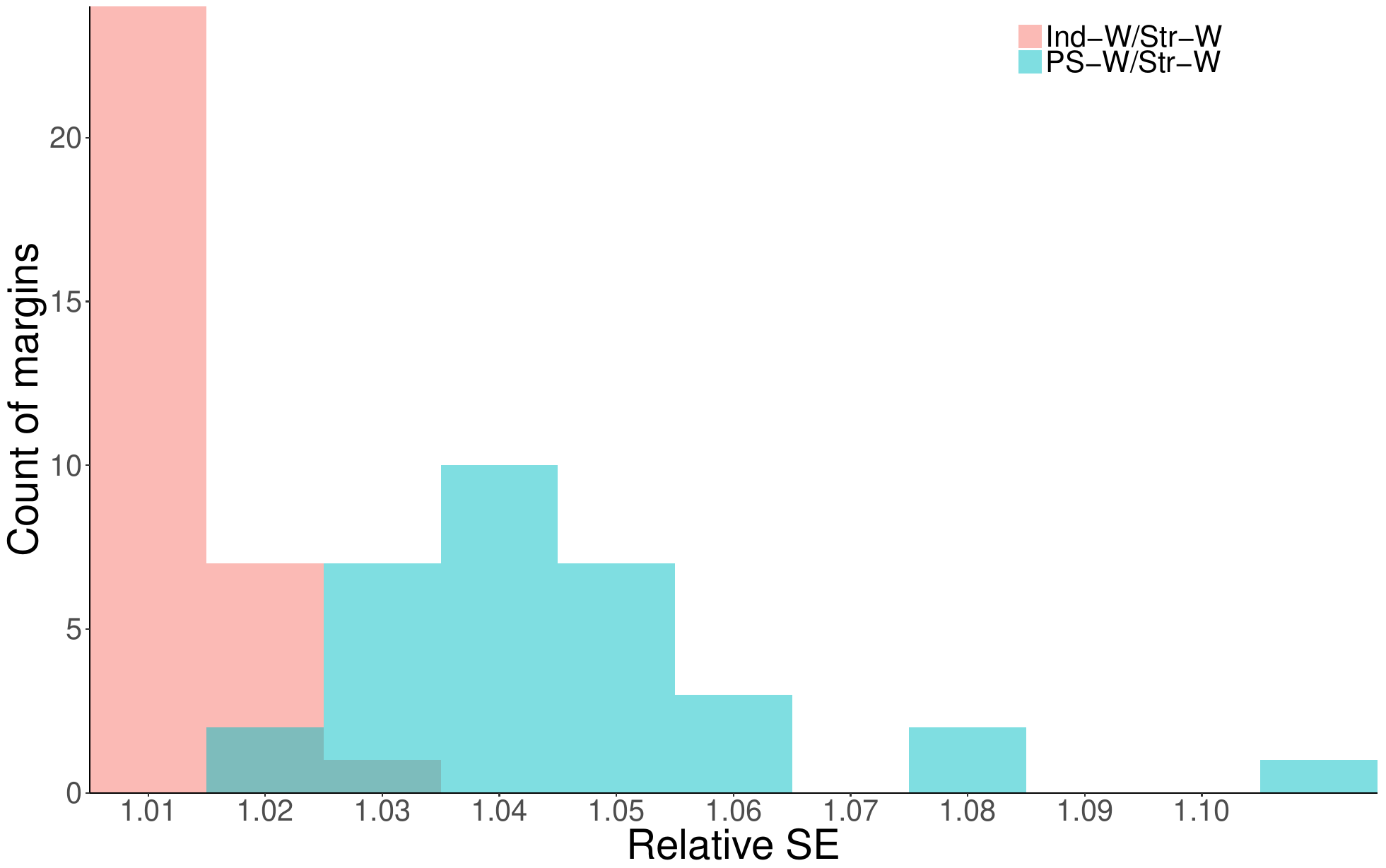}&
\includegraphics[width=.475\textwidth]{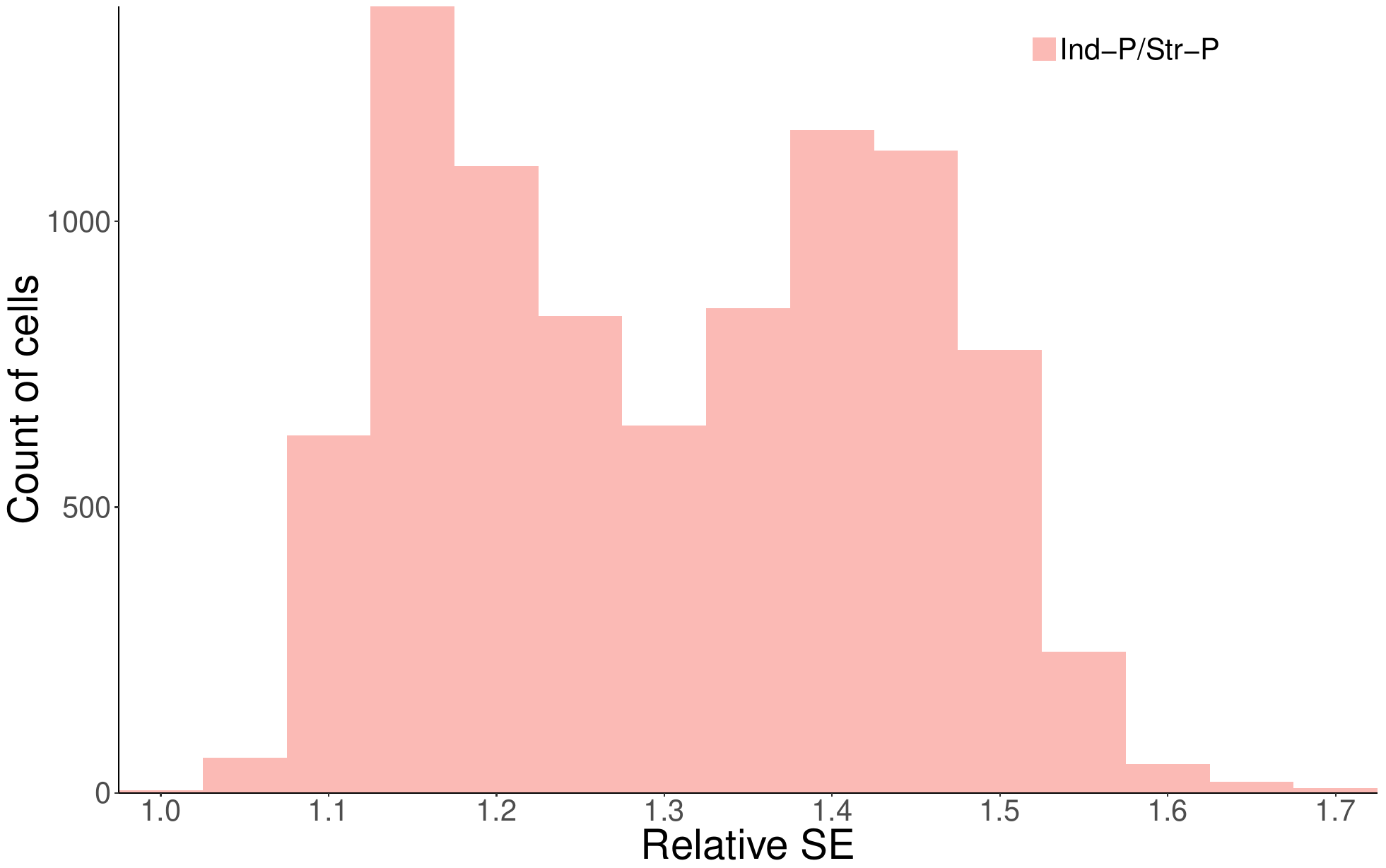}\\
\end{tabular}
\caption{\em Efficiency comparison of prediction and weighting performances on finite population domain inference under a very unbalanced design. The left plot examines the mean estimation across the margins defined by the eight variables. The right plot presents the population cell mean estimation. The model-based weighting and prediction under the structured prior distribution yield smaller SE than those under independent prior. Model-based weighting yields smaller SE than poststratification weighting.}
\label{sim2-se}
\end{figure}

Under the very unbalanced design, the model-based weighting inference under structured prior setting is more efficient than that under independent prior setting or with poststratification weights. We compare the SE of the marginal mean estimates of the eight variables from the three weighting methods and plot the relative ratios in the left plot of Figure~\ref{sim2-se}. The model-based weighting inference has smaller SE than the poststratification weighting, and the weighting under structured prior setting has the smallest SE. Because the sample sizes and the domain sizes are large and the data generation model is sparse, the model-based weighting inference has a little but not much improvement over the poststratification weighting inference due to small smoothing effect. 

The model-based prediction and inference under the structured prior setting are more efficient than that under the independent prior setting. The SEs are smaller with the structured prior than those with the independent prior in the right plot of Figure~\ref{sim2-se}. To demonstrate the efficiency gain, we look at the SEs for the population cell estimates. The Bayesian structural inference generally has smaller variability than that with independent prior, especially in the sparse scenarios.

We assume different outcome and selection models with different covariates with scenarios summarized in Table~\ref{s2-design} and achieve the same evaluation conclusions. 

\section{Application to Longitudinal Study of Wellbeing}
\label{application}

With the background introduced in Section~\ref{problem}, we apply the prediction and weighting inference to the NYC Longitudinal Study of Wellbeing. We match the LSW to the adult population via the ACS. We would like to conduct finite population and domain inference and generate weights allowing for general analysis use. The outcome of interest is the self-reported score of life satisfaction on a 1--10 scale.  We model the outcome as normally distributed, which is not quite correct given that the responses are discrete, but should be fine in practice for the goal of estimating averages. We first include the same eight variables to construct the poststratification cells and use the same estimation model as those in Section~\ref{8var} under the structured prior setting. The posterior inference shows that the variables {\em sex, cldx, eldx}, and {\em psx} are not predictive, and neither are the related high-order interactions. The scale estimates of such terms have posterior median values close to 0 and several large values as long tails. The posterior samples of scales for several high-order interactions among the remaining four variables concentrate around 0, showing these quantities are not predictive. Another complexity is that, for the sample cells of the LSW, the corresponding population cells are not available in the ACS data. This could happen because the sampling frame is not the ACS survey. The population information is unknown for such cells, and untestable assumptions have to be made. The model fitting improves after variable selection when we check the prediction errors for cell estimates.

Hence, we use four variables after selection, {\em age, eth, edu} and {\em pov}, which constructs 500 poststratification cells. The 2002 units in the LSW spread out in 359 cells. The largest sample cell has 86 units, while 92 cells have only one unit. The covariates in the model (\ref{regression}) for cell estimates include the main effects of the four variables, five two-way interactions ({\em age * eth}, {\em age * edu}, {\em eth * edu}, {\em age * inc} and {\em eth * inc}), and two three-way interactions ({\em age * eth * edu} and {\em age * eth * inc}). We implement the fully Bayesian inference with the structured prior distributions. We are interested in estimating the average score of life satisfaction for overall and several subgroups of NYC adults and construct weights for general analysis purposes using the LSW. 

\begin{figure}
\centering
\begin{tabular}{cc}
\includegraphics[width=.475\textwidth]{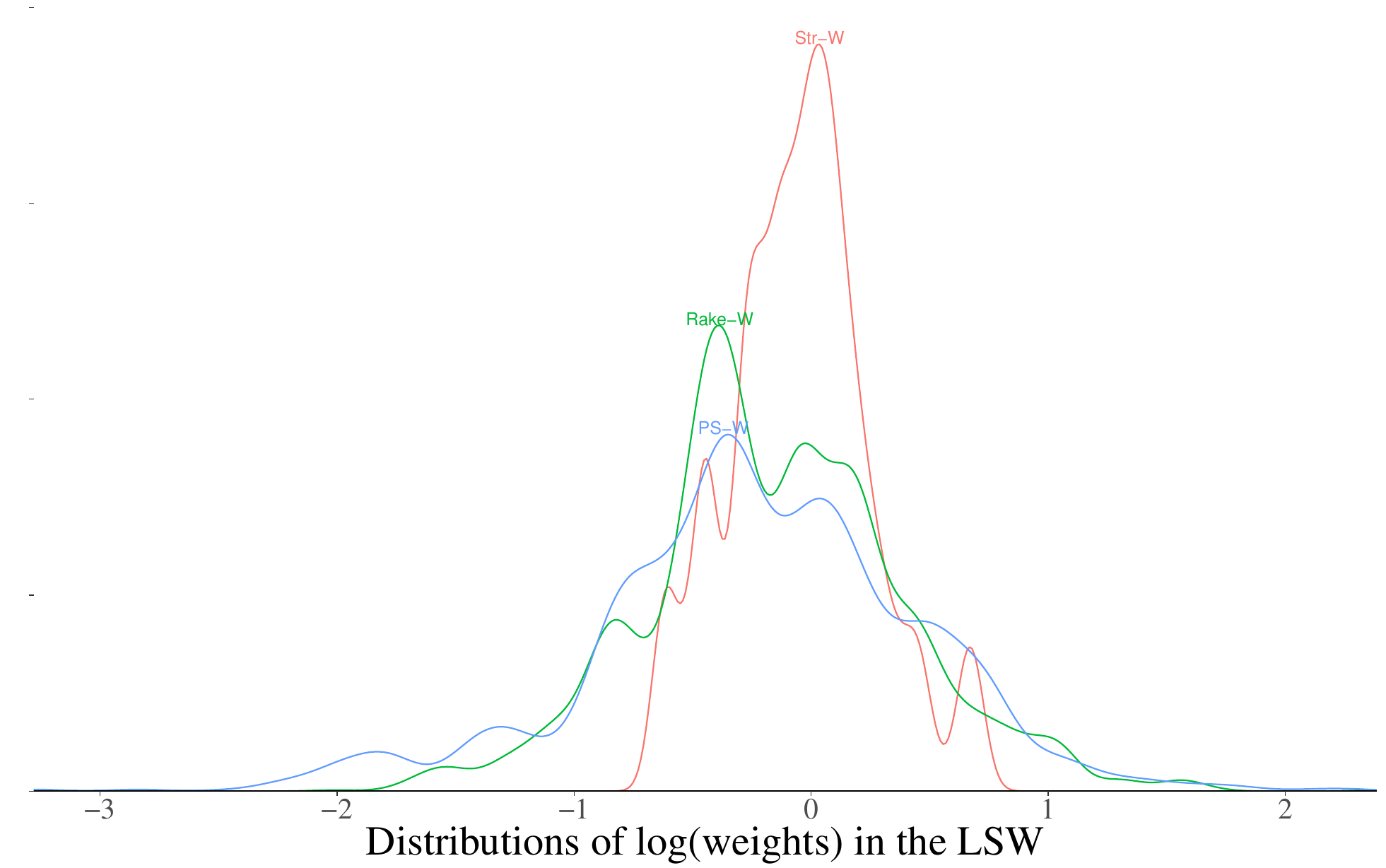}&\includegraphics[width=.475\textwidth]{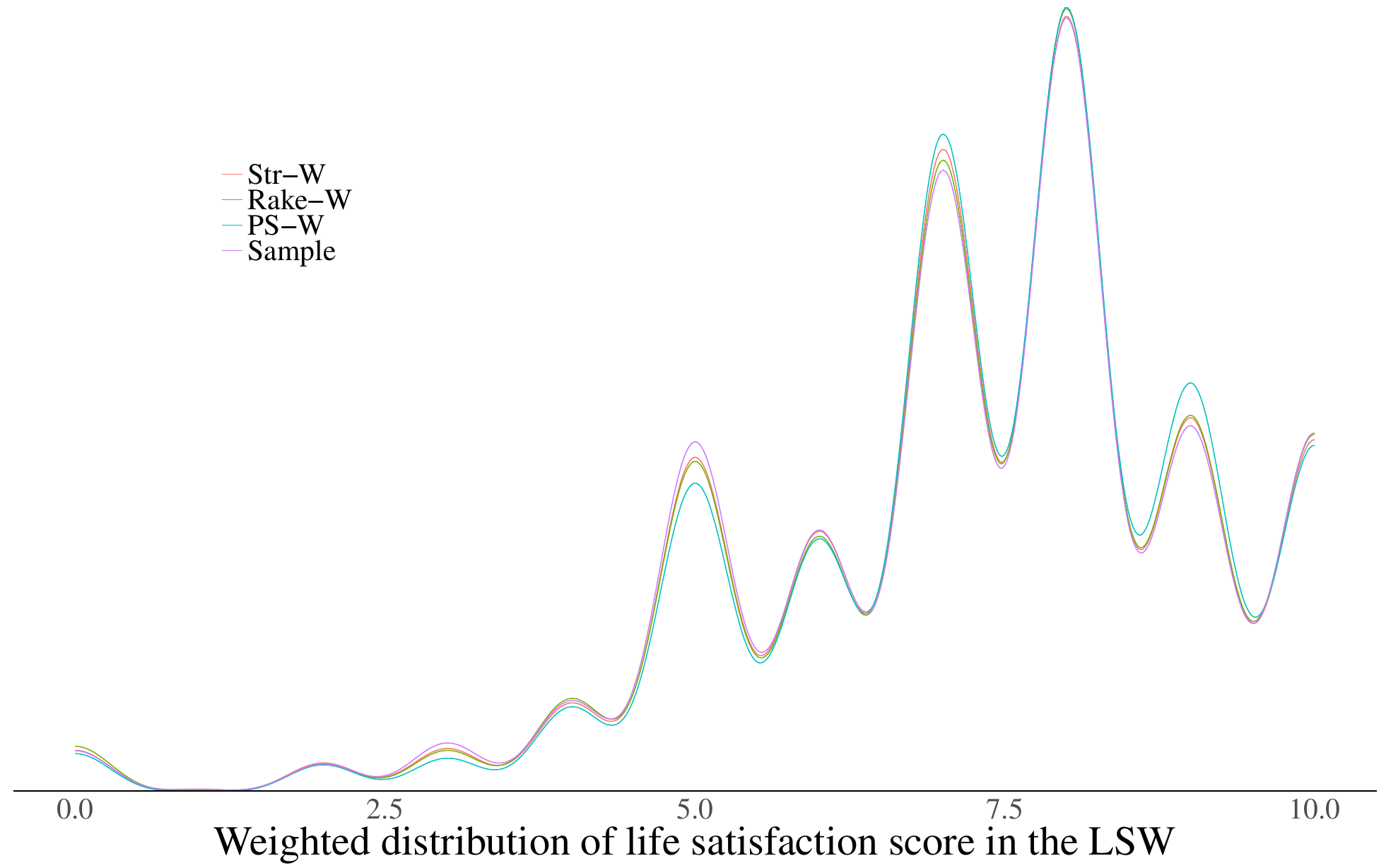}
\end{tabular}
\caption{\em Comparison of generated weights after logarithmic transformation and weighted distributions of life satisfaction score in the LSW. Str-W: model-based weighting under structured prior; Rake-W: weighting via raking adjustment; IP-W: inverse probability of selection weighting, and Sample: sample distribution of the outcome. The weighted distributions are similar between model-based weights and classical weights, but model-based weights are more stable than classical weights.}
\label{lsw-weight}
\end{figure}

The posterior median of the unit scale inside cells $\sigma_y$ is 1.93 with 95\% credible interval $[1.87,1.99]$. The posterior median of the group scale $\sigma_{\theta}$ is 0.79 with 95\% credible interval $[0.65, 1.02]$. These lead to moderately large shrinkage effects between 0.11 and 0.90 with mean 0.30 across cells. The moderate shrinkage effect makes sense based on the four variables and up to three-way interactions being included. The posterior mean values of the model-based weights are presented in the left plot of Figure~\ref{lsw-weight}. We can generate the raking weights after adjustment for the marginal distributions of the four variables and poststratification weights based on the ACS data. The population information is obtained after applying the ACS personal weights.

Comparing with the classical weights, our model-based weights have smaller variability with standard deviation 0.32 and the ratio of the maximum and minimum value 3.87, and these values are much smaller than those for the raking and poststratification weights, as shown in Table~\ref{lsw-est}. The right plot in Figure~\ref{lsw-weight} shows the distribution of the lift satisfaction score after weighting. The model-based weighted distributions and classically weighted distributions are similar as expected, which is consistent with the results in Section~\ref{8var}. The weighting process adjusts for the sample distribution by upweighting the high scores and downweighting the low scores. The LSW oversamples poor residents who tend not be satisfied with life, and the weighting adjustment balances the discrepancy.

\begin{table}
\centering
\small
\begin{tabular}{lllll}
& Str-P & Str-W & Rake-W & PS-W \\ 
  \hline
SD of weights / mean of weights & & 0.32 & 0.66 & 0.80 \\
Max weight / min weight &  & 3.87 & 81.28 & 274.65 \\ 
  \hline
 \multicolumn{5}{l}{Overall average for NYC adults ($n=2002$)}\\
  Est & 7.24 & 7.23 & 7.24 & 7.30 \\ 
  SE & 0.05 & 0.05 & 0.05 & 0.06\\ 
  \hline
 \multicolumn{5}{l}{\begin{tabular}{@{}l@{}}Average for middle-aged, college-educated whites with poverty gap $> 300\%$ ($n=222$)\end{tabular}}\\
Est & 7.40 & 7.34 & 7.34 & 7.34\\ 
SE & 0.10 & 0.11 & 0.11 & 0.11 \\ 
 \hline
\multicolumn{5}{l}{\begin{tabular}{@{}l@{}}Average for elders with poverty gap $< 200\%$ ($n=154$)\end{tabular}}\\
Est & 7.37 & 7.52 & 7.49 & 7.53 \\ 
  SE & 0.15 & 0.18 & 0.19 & 0.22 \\ 
\hline
 \multicolumn{5}{l}{\begin{tabular}{@{}l@{}}Average for blacks with poverty gap $< 50\%$ ($n=57$)\end{tabular}}\\
Est & 7.01 & 7.16 & 7.30 & 7.16 \\ 
SE & 0.18 & 0.26 & 0.28 & 0.29
\end{tabular}
\caption{\em Comparison of prediction and weighting performances on estimating various domain averages for life satisfaction in the LSW. Str-P: model-based prediction under the structured prior; Str-W: model-based weighting under structured prior; Rake-W: weighting via raking adjustment; and PS-W: poststratification weighting. }
\label{lsw-est}
\end{table}

\begin{figure}
\centering
\begin{tabular}{cc}
\includegraphics[width=.475\textwidth]{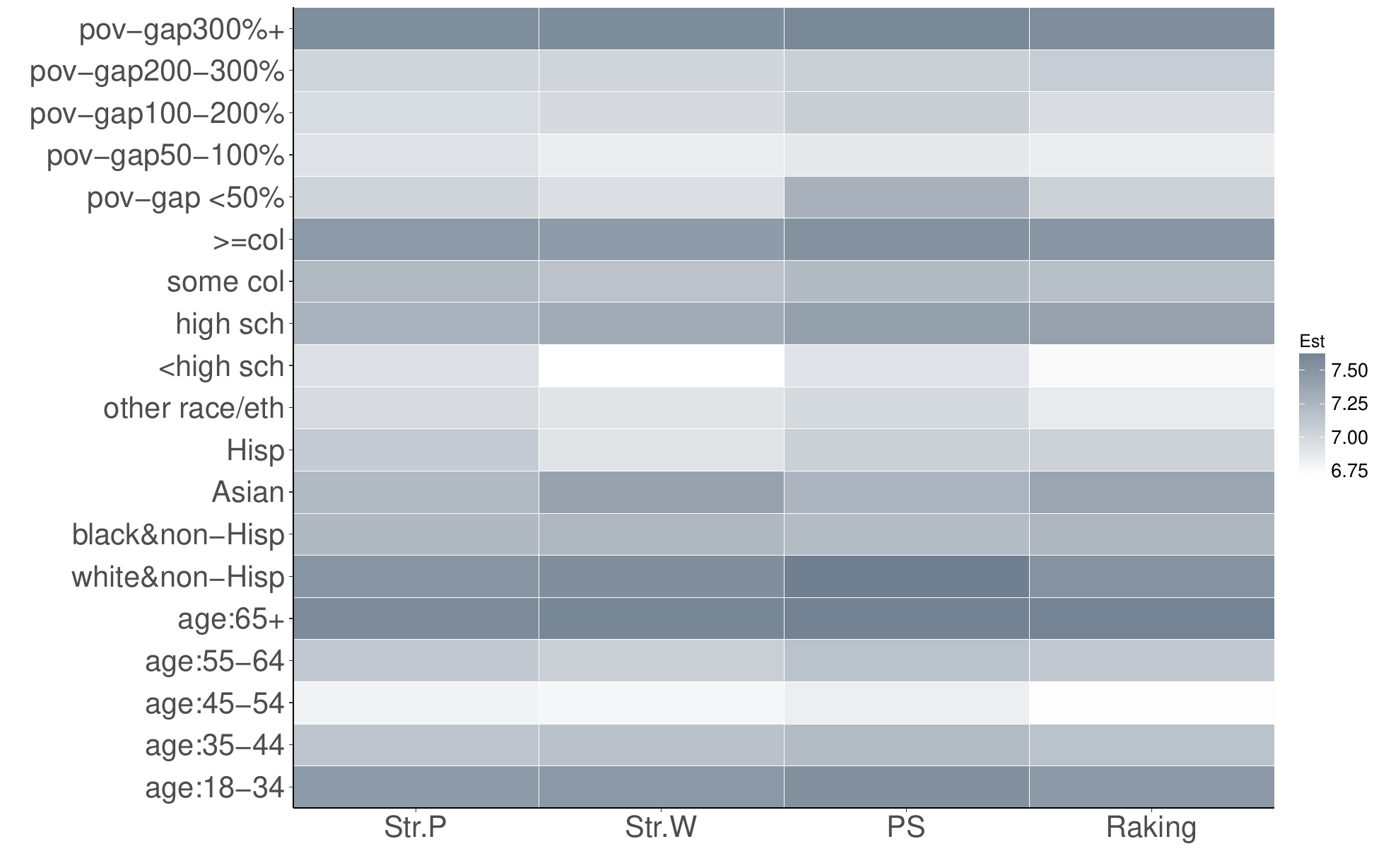}&\includegraphics[width=.475\textwidth]{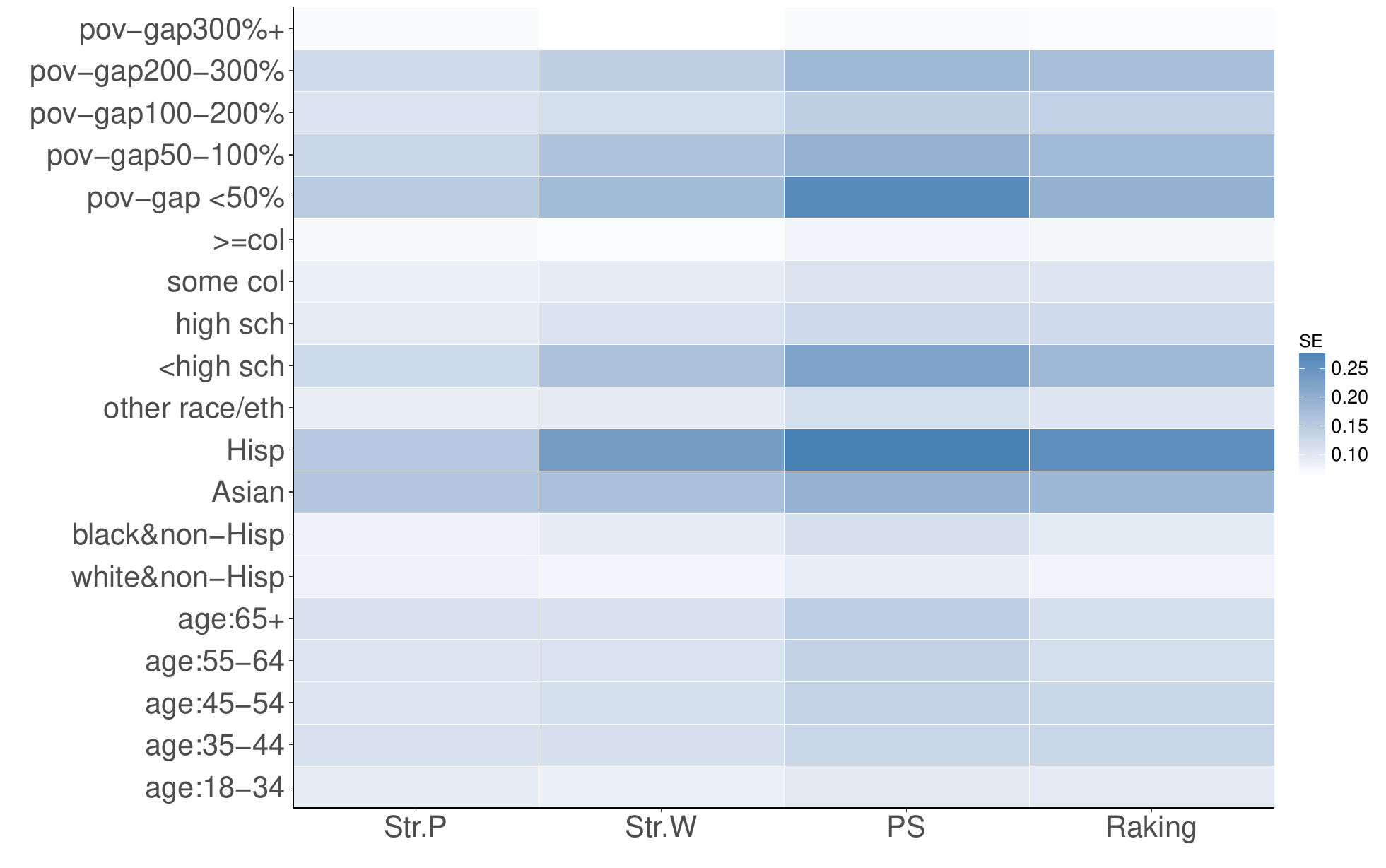}
\end{tabular}
\caption{\em Comparison of predictions and weighting performances on estimating life satisfaction score across the margins of four variables in the LSW. Str-P: model-based prediction under the structured prior; Str-W: model-based weighting under structured prior; Rake-W: weighting via raking adjustment; and PS-W: poststratification weighting. Model-based predictions and weighting generate different estimates for several subsets and are generally more efficient comparing with classical weighting.}
\label{lsw-mar}
\end{figure}

Table~\ref{lsw-est} and Figure~\ref{lsw-mar} present the finite population and domain inference. The average score of life satisfaction for NYC adults is 7.24 with standard error 0.05, predicted by the structural model. The estimate is similar to that under model-based weighting and raking inferences, but lower than the poststratification weighting inference. However, the difference is not significant. For example, the structural model predicts the average score of life satisfaction for middle-aged, college-educated whites with income more than three times the poverty level as
 7.40 with standard error 0.10, higher than that under weighting inferences. Nevertheless, the predicted scores for the elder with relatively low income (7.37 with SE 0.15) and low-income black New Yorkers (7.01 with SE 0.18) are lower than those under weighting inferences. The discrepancy could be explained by the nonrepresentativeness of the LSW and the deep interactions included by the model. The subgroup of individuals who are middle-aged, college-educated whites may be undercovered in the LSW---as empty poststratification cells occurring---with overcoverage among elderly poor blacks. Weighting the collected samples cannot infer or extrapolate inference on those who are not present in the survey. Though the differences are not significant, inferences conditioning on the collected samples cannot recover the truth, especially for the empty cell estimates. Figure~\ref{lsw-mar} shows the model-based prediction yields a higher score for young, highly educated and Hispanic NYC adults, but a lower score for those with poverty gap $<50\%$, comparing with the weighted inference.

The SEs are similar for the overall mean estimation between predictions and various weighting inferences because of the large sample size. For domain estimation, the model-based prediction and weighting are more efficient than that with raking and poststratification weighting, and the model-based prediction has the smallest standard error. The efficiency gains of model-based prediction and weighting are further demonstrated by domain mean estimation for life satisfaction scores across the marginal levels of four variables, shown in Figure~\ref{lsw-mar}. The model-based prediction and weighting particularly improve small domain estimation and increase the efficiency. 

Survey practitioners often compare the weighted distribution of socio-demographics with the population distribution to check the weighting. While weighting diagnostics need further research and management, we follow this routine to compare the model-based and classical weights. We calculate the Euclidean distances between the weighted distributions and the population distribution for the main effects and high-order interactions among the four variables in the LSW, shown in Table~\ref{prob-dist} in Appendix~\ref{appendix}. The weighted distributions are generally close to the true distributions. Raking focuses on adjusting for the marginal distributions of calibration variables but distorts the joint distributions, where the dependency structure is determined only by the sample without calibration. The poststratification weighting adjusts for the joint distribution, but empty cells in the sample present from the exact matching. The unbalanced cell structure yields unstable inference. The model-based weighting smooths the poststratification weightings and outperforms raking to match the distributions of three-way and four-way interaction terms. Practitioners often rely upon the marginal distributions to evaluate weighting performances, thus in favor of raking. However, raking yields high variable and potentially biased inferences, shown in the Section~\ref{simulation}, even in the cases when raking adjustment is correct.  Modification of model-based weighting to satisfy such desire on matching marginal distributions will be a future extension to incorporate constraints.

\section{Discussion}
\label{discussion}

We combine Bayesian prediction and weighting as a unified approach to survey inference. Multilevel regression with structured prior distributions and poststratification on the population inference yield efficient estimation when accounting for the design feature. The computation is implemented via Stan and disseminated through the R package {\tt rstanarm} for public use, and the software development promotes the model-based approaches in survey research and operational practice. We construct stable and calibrated model-based weights to solve the problems of classical weights. This article builds up the model-based prediction and weighting framework and serves as the first contribution to evaluate the statistical properties of model-based weights and compare the performances with classical weighting. Model-based weights are smoothed across poststratification cells and improve small domain estimation. 

The structured prior uses the hierarchical structure between the main effects and high-order interaction terms to introduce multiplicative constraints on the corresponding scale parameters and informs variable selection. Model improvement can be done after post-processing the posterior inferences. The Bayesian structural model yields more stable inference than that with independent prior distributions. Such hierarchy assumption may not be valid for special cases, such as the Exclusive-Or problem where two variables show no main effects but a perfect interaction. However, we do not have strong evidence in the application studies against the plausibility of hierarchy. Furthermore, the unified prediction and weighting approach is well equipped to deal with complex survey designs and big data in surveys, such as streaming data and combining multiple survey studies. 

The general MRP framework is open to flexible modeling strategies. In this article, we illustrate by a regression model with all variables of interest and the high-order interactions and incorporate structured prior distributions for regularization. Other approaches, such as nonparametric models and machine learning tools, can be implemented under the MRP framework, being robust against model misspecification. \cite{bnfp:ba15} use Gaussian process regression models to borrow information across poststratification cells based on the distances between the inverse inclusion probability weights. Further extensions include applying such flexible approaches to weight smoothing and deriving the model-based weights. 

The broad application opportunities come with various challenges that need further investigation. The model-based weights are outcome dependent, which improves the efficiency but potentially reduces the robustness. Survey organizers prefer a set of weights that can be used for general analysis purpose, without being sensitive to outcome selection. We can compare different weights constructed by several important outcomes and conduct sensitivity analysis. When the model-based weights give different inference conclusions, we recommend choosing the set of weights that generate the most reasonable results, with scientific reasoning and be consistent with the population inference. 

The weighted marginal distributions of the calibration variables are a bit different from the population inferences, as in Section~\ref{application}, which does not meet the usual weighting diagnosis standard of survey organizers. The model-weights tend to match the joint distribution to that in the population, but weight smoothing may bring in bias. Tradeoff constraints can be induced to the model to match the marginal distributions. 

Another practical challenge is that the population distribution of the calibration variables may be unknown, that is, the population poststratification cell sizes $N_j$'s are unknown. A supplemental model is needed to allow estimation of this information from the sample and integrated with MRP to propagate all sources of uncertainty as an extension, similar to the framework in~\cite{BayesRake18} by incorporating known margins. The model-based predictions and weighting inferences need further extensions to handle discrete outcomes, inference on regression coefficients and non-probability or informative sampling designs~\citep{Kim:Skinner:BM13}. It will be useful to link these ideas on survey inference with the biostatistical and econometrics literature on inverse propensity score and doubly robust weighting~\citep{kang:schafer07}.

\section*{Acknowledgements}
We thank the National Science Foundation, National Institutes of Health, Office of Naval Research, Institute of Education Sciences, and Sloan Foundation for grant support.

\pagebreak

\appendix

\section{Example code}
\label{code}
Here we present code for the application described in the data. We have written a
function \texttt{model\_based\_cell\_weights} to calculate the model-based weights
from a fitted {\tt rstanarm} model. 
\begin{small}
\begin{verbatim}
model_based_cell_weights <- function(object, cell_table) {
  stopifnot(
    is.data.frame(cell_table),
    colnames(cell_table) == c("N", "n")
  )
  draws <- as.matrix(object)
  Sigma <- draws[, grep("^Sigma\\[", colnames(draws)), drop = FALSE]
  sigma_theta_sq <- rowSums(Sigma)
  sigma_y_sq <- draws[, "sigma"]^2
  Ns <- cell_table[["N"]]  # population cell counts
  ns <- cell_table[["n"]]  # sample cell counts
  J <- nrow(cell_table)
  N <- sum(Ns)
  n <- sum(ns)
  # implementing equation 7 in the paper (although i did some algebra first to 
  # simplify the expression a bit)
  Nsy2 <- N * sigma_y_sq
  ww <- matrix(NA, nrow = nrow(draws), ncol = J)
  for (j in 1:J) {
    ww[, j] <- 
      (Nsy2 + n * Ns[j] * sigma_theta_sq) / (Nsy2 + N * ns[j] * sigma_theta_sq)
  }
  return(ww)
}
# prepare population data: acs_ad has age, eth, edu and inc
acs_ad %>% 
  mutate(
    cell_id =  paste0(age, eth, edu, inc)
  ) -> acs_ad
acs_design <- svydesign(id = ~1, weights = ~perwt, data = acs_ad)
agg_pop <- 
  svytable( ~ age + eth + edu + inc, acs_design) %>% 
  as.data.frame() %>%
  rename(N = Freq) %>%
  mutate(
    cell_id = paste0(age, eth, edu, inc) 
  ) %>%
  filter(cell_id %in% acs_ad$cell_id)
# prepare data to pass to rstanarm
# SURVEYdata has 4 variables used for weighting: age, eth, edu and inc; and outcome Y
dat_rstanarm <-
  SURVEYdata %>%
  mutate(
    cell_id = paste0(age, eth, edu, inc)
  )%>% 
  group_by(age, eth, edu, inc) %>%
  summarise(
    sd_cell = sd(Y),
    n = n(),
    Y = mean(Y),
    cell_id = first(cell_id)
  ) %>%
  mutate(sd_cell = if_else(is.na(sd_cell), 0, sd_cell)) %>%
  left_join(agg_pop[, c("cell_id", "N")], by = "cell_id")
# Stan fitting under structured prior in rstanarm
fit <-
  stan_glmer(
    formula = 
      Y ~ 1 + (1 | age) + (1 | eth) + (1 | edu) + (1 | inc) +
      (1 | age:eth) + (1 | age:edu) + (1 | age:inc) +
      (1 | eth:edu) + (1 | eth:inc) + 
      (1 | age:eth:edu) + (1 | age:eth:inc),
    data = dat_rstanarm,  iter = 1000, chains = 4, cores = 4,
    prior_covariance = 
      rstanarm::mrp_structured(
        cell_size = dat_rstanarm$n, 
        cell_sd = dat_rstanarm$sd_cell, 
        group_level_scale = 1,
        group_level_df = 1
      ),
    seed = 123,
    prior_aux = cauchy(0, 5),
    prior_intercept = normal(0, 100, autoscale = FALSE), 
    adapt_delta = 0.99
  )
# model-based weighting
cell_table <- fit$data[,c("N","n")]
weights <- model_based_cell_weights(fit, cell_table)
weights <- data.frame(w_unit = colMeans(weights),
                      cell_id = fit$data[["cell_id"]],
                      Y = fit$data[["Y"]],
                      n = fit$data[["n"]]) %>%
           mutate(
             w = w_unit / sum(n / sum(n) * w_unit), # model-based weights
             Y_w = Y * w
           ) 
with(weights, sum(n * Y_w / sum(n)))# mean estimate
\end{verbatim}
\end{small}

\pagebreak
\section{Simulation designs}
\label{appendix}

Here we present the simulation designs, coefficient values, and comparison on the weighted distributions of socio-demographics as a supplement to Sections \ref{simulation} and \ref{application}.

\begin{table}
\begin{center}
\small
\begin{tabular}{c|cc|cc|cc|cc|cc|cc|cc}
&\multicolumn{2}{|c|}{Case 1}&\multicolumn{2}{|c|}{Case 2}&\multicolumn{2}{|c|}{Case 3}&\multicolumn{2}{|c|}{Case 4}&\multicolumn{2}{|c|}{Case 5}&\multicolumn{2}{|c}{Case 6}&\multicolumn{2}{|c}{Case 7}\\
\hline
&O&S&O&S&O&S&O&S&O&S&O&S&O&S\\
age&$\checkmark$&$\checkmark$&$\checkmark$&$\checkmark$&$\checkmark$&$\checkmark$&$\checkmark$&$\checkmark$&$\checkmark$&$\checkmark$&$\checkmark$&$\checkmark$&$\checkmark$&$\checkmark$\\
eth&$\checkmark$&$\checkmark$&$\checkmark$&$\checkmark$&$\checkmark$&$\checkmark$&$\checkmark$&$\checkmark$&$\checkmark$&\checkmark&&$\checkmark$&&$\checkmark$\\
edu&$\checkmark$&$\checkmark$&$\checkmark$&$\checkmark$&$\checkmark$&&$\checkmark$&$\checkmark$&$\checkmark$&&$\checkmark$&$\checkmark$&$\checkmark$&\\
age*eth&$\checkmark$&&&$\checkmark$&$\checkmark$&$\checkmark$&&&&$\checkmark$&&&&$\checkmark$\\
age*edu&$\checkmark$&&&$\checkmark$&$\checkmark$&&&&&&$\checkmark$&&$\checkmark$&\\
eth*edu&$\checkmark$&&&$\checkmark$&$\checkmark$&&&&&&&&&\\
age*eth*edu&$\checkmark$&&&$\checkmark$&$\checkmark$&&&&&&&&&
\end{tabular}
\end{center}
\caption{\em Covariates in the outcome (O) and selection (S) models for slightly unbalanced design.}
\label{s1-design}
\end{table}

\begin{table}[h!]
\begin{center}
\small
\begin{tabular}{c|p{4cm}|l|p{4cm}}
&All & Main effects & Two variables\\
\hline
age&(0.5 1.375 2.25 3.125 4)&(0.5 1.375 2.25 3.125 4)&(0.5 1.375 2.25 3.125 4)\\
eth&(-2 -1 0 1 2)&(2 -1 0 1 2)& $\vec{0}$\\
edu&(3 2 1 0)&(3 2 1 0)&(3 2 1 0)\\
age*eth&(4 2 1 1 3 3 2 1 1 1 2 3 2 2 1 4 4 3 2 3 2 4 1 4 1)& $\vec{0}$& $\vec{0}$\\
age*edu&(-2 -1 2 2 1 -2 2 1 0 -2 1 -2 -1 2 1 -1 -1 2 0 2)& $\vec{0}$&(2 0 -2 -2 1 1 -1 -2 -2 -1 -1 1 0 -1 -1 2 2 1 -1 0 )\\
eth*edu&(1 -2 0 -3 -1 0 -1 -2 0 -1 -3 -3 0 -1 -1 0 0 -1 0 -1)& $\vec{0}$& $\vec{0}$\\
age*eth*edu&(-1 -0.5 0.5 -1 -1 -0.5 -1 0 -1 0 -1 0 1 1 0.5 1 1 -1 -1 0 -1 -0.5 -0.5 -1 1 -1 -0.5 -1 1 0 0.5 0.5 1 0.5 1 1 1 0.5 1 0 0 -0.5 0 1 -1 -1 0 -1 -1 -1 -0.5 -0.5 0 1 -1 0 0 -0.5 1 -0.5 0.5 -1 1 0 1 0 -1 0 -0.5 1 -0.5 -1 -0.5 0 0.5 -0.5 1 0.5 -0.5 0.5 0 1 0 1 0.5 0.5 0.5 0 0 -0.5 1 -1 0 1 1 1 1 -0.5 -1 -1)& $\vec{0}$& $\vec{0}$
\end{tabular}
\end{center}
\caption{\em Assumed regression coefficients in the {\em outcome} model for the simulation using a slightly unbalanced design.}
\label{s1-response-coef}
\end{table}

\begin{table}
\begin{center}
\small
\begin{tabular}{c|p{4cm}|l|p{4cm}}
&All&Main effects & Two variables\\
\hline
Intercept&-2&-2&-2\\
age&(-2 -1.75 -1.5 -1.25 -1)&(0 0.5 1 1.5 2)&(-2 -1.5 -1 -0.5 0)\\
eth&(-1 -0.25 0.5 1.25 2)&(-2 -1.5 -1 -0.5 0)&(-1 -0.5 0 0.5 1)\\
edu&(0 0.67 1.33 2)&(0 1 2 3)& $\vec{0}$\\
age$\times$eth&(1 1 -1 1 -1 1 -1 0 0 -1 0 0 -1 1 0 0 -1 1 1 -1 -1 0 1 -1 1)& $\vec{0}$&(-1 1 1 1 -1 -1 -1 0 -1 -1 -1 -1 1 -1 -1 0 1 1 -1 1 -1 -1 1 0 0)\\
age$\times$edu&(0 1 -1 -1 0 1 1 0 1 0 1 -1 -1 1 1 -1 0 -1 1 1 )& $\vec{0}$& $\vec{0}$\\
eth$\times$edu&(-1 -1 0 -1 -1 1 1 1 1 0 -1 0 -1 0 -1 1 0 -1 -1 -1 )& $\vec{0}$& $\vec{0}$\\
age$\times$eth$\times$edu& (0.8 -0.4 0.6 -0.2 0.8 0.2 0.4 0.8 0.4 -0.6 -0.8 -0.4 -0.8 -0.4 0.4 -1 0.6 -0.8 -0.6 0.6 -0.2 0.2 0.6 -0.6 0 0 -1 -0.2 0.6 0.8 -0.4 0.2 -0.8 0.4 0.6 -0.6 0.8 0 0.2 -1 1 0.4 0 0.8 -0.2 0 0 0.6 -0.8 -0.8 -0.2 0.4 -1 -0.8 1 -0.2 0 0.8 0.6 0.8 -0.2 -0.2 -0.8 1 0.8 0.8 -0.4 -0.8 0.4 -0.4 1 -0.6 -1 -0.6 -0.2 1 1 -0.2 1 0.6 0.4 0.8 0.2 -0.2 -0.6 0 0.8 -0.4 0.4 0.4 0.6 -1 -0.8 -0.8 1 1 0.4 0.6 0.4 0.8)&$\vec{0}$& $\vec{0}$
\end{tabular}
\end{center}
\caption{\em Assumed regression coefficients in the {\em selection}  model for the simulation using a slightly unbalanced design.}
\label{s1-selection-coef}
\end{table}

\begin{table}
\begin{center}
\small
\begin{tabular}{c|cc|cc|cc|cc}
&\multicolumn{2}{|c|}{Case 1}&\multicolumn{2}{|c|}{Case 2}&\multicolumn{2}{|c|}{Case 3}&\multicolumn{2}{c}{Case 4}\\
\hline
&O&S&O&S&O&S&O&S\\
age&$\checkmark$&$\checkmark$&$\checkmark$&$\checkmark$&$\checkmark$&$\checkmark$&$\checkmark$&$\checkmark$\\
eth&$\checkmark$&$\checkmark$&$\checkmark$&$\checkmark$&$\checkmark$&$\checkmark$&$\checkmark$&$\checkmark$\\
edu&$\checkmark$&$\checkmark$&$\checkmark$&$\checkmark$&$\checkmark$&$\checkmark$&$\checkmark$&$\checkmark$\\
sex&$\checkmark$&$\checkmark$&$\checkmark$&$\checkmark$&$\checkmark$&$\checkmark$&$\checkmark$&$\checkmark$\\
pov&$\checkmark$&$\checkmark$&$\checkmark$&$\checkmark$&$\checkmark$&$\checkmark$&$\checkmark$&$\checkmark$\\
cld&&$\checkmark$&&$\checkmark$&&$\checkmark$&$\checkmark$&$\checkmark$\\
eld&$\checkmark$&$\checkmark$&&$\checkmark$&$\checkmark$&$\checkmark$&$\checkmark$&$\checkmark$\\
fam&$\checkmark$&$\checkmark$&&$\checkmark$&$\checkmark$&$\checkmark$&$\checkmark$&$\checkmark$\\
age*eth&$\checkmark$&$\checkmark$&&&$\checkmark$&&&$\checkmark$\\
age*edu&$\checkmark$&$\checkmark$&&&$\checkmark$&&&$\checkmark$\\
eth*edu&$\checkmark$&$\checkmark$&&&$\checkmark$&& &$\checkmark$\\
eth*pov&$\checkmark$&$\checkmark$&&&$\checkmark$&&&$\checkmark$\\
age*pov&$\checkmark$&$\checkmark$&&&$\checkmark$&&&$\checkmark$\\
pov*fam&$\checkmark$&$\checkmark$&&&$\checkmark$&&&$\checkmark$\\
pov*eld&$\checkmark$&$\checkmark$&&&$\checkmark$&&&$\checkmark$\\
pov*cld&&$\checkmark$&&&&&&$\checkmark$\\
age*eth*edu&$\checkmark$&$\checkmark$&&&$\checkmark$&&&$\checkmark$\\
age*eth*pov&$\checkmark$&$\checkmark$&&&$\checkmark$&&&$\checkmark$
\end{tabular}
\end{center}
\caption{\em Covariates in the outcome (O) and selection (S) models for a very unbalanced design.}
\label{s2-design}
\end{table}

\begin{table}
\begin{center}
\small
\begin{tabular}{c|p{7cm}|p{7cm}}
&O&S\\
\hline
age&(2 0 -2 -2 1)&(0 0.75 1.5 2.25 3)\\
eth&(1 -1 -2 -2 -1)&(-1 -0.5 0 0.5 1)\\
edu&(-1 1 0 -1)&(0 0.67 1.33 2)\\
sex&(-1 2)&(-1 0)\\
pov&(2 1 -1 0 -1)&(0 1 2 3 4)\\
cld&$\vec{0}$&(-1 -0.33 0.33 1)\\
eld&$\vec{0}$&(-2 -1 0)\\
fam&$\vec{0}$&(-1 -0.67 -0.33 0)
\end{tabular}
\end{center}
\caption{\em Assumed regression coefficients in the outcome (O) and selection (S) models for a very unbalanced design.}
\label{s2-design-case2}
\end{table}

\begin{table}
\centering
\small
\begin{tabular}{lrrr}
 & Str-W & PS-W & Rake-W \\ 
 \hline
 {\em age} & 0.04 & 0.02 & 0.00\\
 {\em eth}& 0.08 & 0.06 & 0.00 \\ 
{\em edu} & 0.08 & 0.03 & 0.00 \\ 
{\em inc}& 0.02 & 0.02 & 0.00\\ 

{\em age * eth}& 0.05 & 0.03 & 0.05 \\
{\em age * edu}& 0.05 & 0.02 & 0.05 \\
{\em age * inc}& 0.03 & 0.01 & 0.03 \\

{\em eth * edu} & 0.06 & 0.04 & 0.05 \\ 
{\em eth * inc}& 0.04 & 0.04 & 0.03 \\ 
{\em edu * inc} & 0.06 & 0.03 & 0.04 \\ 

{\em age * eth * edu}& 0.03 & 0.02 & 0.05 \\ 
{\em age * eth * inc}& 0.03 & 0.02 & 0.04 \\
{\em age * edu * inc } & 0.03 & 0.01 & 0.04 \\
{\em eth * edu * inc} &0.04 & 0.02 & 0.04 \\
{\em age * eth * edu * inc}& 0.02 & 0.01& 0.04
\end{tabular}
\caption{\em Euclidean distances between the weighted distributions and the population distribution. Str-W: model-based weighting under structured prior; Rake-W: weighting via raking adjustment; and PS-W: poststratification weighting. }
\label{prob-dist}
\end{table}

\pagebreak

\newpage
\bibliographystyle{chicago}
\bibliography{/Users/yajuan/Box/bibs/weighting-2020}

\begin{thebibliography}{}

\bibitem[\protect\citeauthoryear{{ACS Weighting Method}}{{ACS Weighting
  Method}}{2014}]{acsweighting2014}
{ACS Weighting Method} (2014).
\newblock {\em American Community Survey Design and Methodology, Chapter 11:
  Weighting and Estimation}.
\newblock United States Census Bureau.

\bibitem[\protect\citeauthoryear{Beaumont}{Beaumont}{2008}]{beaumont08}
Beaumont, J.~P. (2008).
\newblock A new approach to weighting and inference in sample surveys.
\newblock {\em Biometrika\/}~{\em 95}, 539--553.

\bibitem[\protect\citeauthoryear{Breidt and Opsomer}{Breidt and
  Opsomer}{2017}]{model-assist-review-SS17}
Breidt, F. and J.~Opsomer (2017).
\newblock Model-assisted survey estimation with modern prediction techniques.
\newblock {\em Statistical Science\/}~{\em 32}, 190--205.

\bibitem[\protect\citeauthoryear{Breidt}{Breidt}{2008}]{Breidt08}
Breidt, F.~J. (2008).
\newblock Endogenous post-stratification in surveys: {C}lassifying with a
  sample-fitted model.
\newblock {\em Annals of Statistics\/}~{\em 36}, 403--427.

\bibitem[\protect\citeauthoryear{Carvalho, Polson, and Scott}{Carvalho
  et~al.}{2010}]{horseshoe10}
Carvalho, C.~M., N.~G. Polson, and J.~G. Scott (2010).
\newblock The horseshoe estimator for sparse signals.
\newblock {\em Biometrika\/}~{\em 97}, 465--480.

\bibitem[\protect\citeauthoryear{Chambers, Dorfman, and Wehrly}{Chambers
  et~al.}{1993}]{robustblup:Chambers:JASA93}
Chambers, R.~L., A.~H. Dorfman, and T.~E. Wehrly (1993).
\newblock Bias robust estimation in finite populations using nonparametric
  calibration.
\newblock {\em Journal of the American Statistical Association\/}~{\em 88},
  260--269.

\bibitem[\protect\citeauthoryear{Chen, Elliott, Haziza, Yang, Ghosh, Little,
  Sefransk, and Thompson}{Chen et~al.}{2017}]{samsi:review17}
Chen, Q., M.~R. Elliott, D.~Haziza, Y.~Yang, M.~Ghosh, R.~Little, J.~Sefransk,
  and M.~Thompson (2017).
\newblock Approaches to improving survey-weighted estimates.
\newblock {\em Statistical Science\/}~{\em 32\/}(2), 227--248.

\bibitem[\protect\citeauthoryear{Dahlke, Breidt, Opsomer, and Keilegom}{Dahlke
  et~al.}{2013}]{Dahlke13}
Dahlke, M., F.~Breidt, J.~Opsomer, and I.~V. Keilegom (2013).
\newblock Nonparametric endogenous post-stratification in surveys.
\newblock {\em Statistica Sinica\/}~{\em 23}, 189--211.

\bibitem[\protect\citeauthoryear{Deville and S{\"a}rndal}{Deville and
  S{\"a}rndal}{1992}]{greg92}
Deville, J.~C. and C.~E. S{\"a}rndal (1992).
\newblock Calibration estimators in survey sampling.
\newblock {\em Journal of the American Statistical Association\/}~{\em 87},
  376--382.

\bibitem[\protect\citeauthoryear{Deville, Sarndal, and Sautory}{Deville
  et~al.}{1993}]{gr-rake93}
Deville, J.-C., C.-E. Sarndal, and O.~Sautory (1993).
\newblock Generalized raking procedures in survey sampling.
\newblock {\em Journal of the American Statistical Association\/}~{\em
  88\/}(423), 1013--1020.

\bibitem[\protect\citeauthoryear{Elliott}{Elliott}{2007}]{elliott07}
Elliott, M.~R. (2007).
\newblock Bayesian weight trimming for generalized linear regression models.
\newblock {\em Journal of Official Statistics\/}~{\em 33\/}(1), 23--34.

\bibitem[\protect\citeauthoryear{Elliott and Little}{Elliott and
  Little}{2000}]{modeltrim-elliottandlittle00}
Elliott, M.~R. and R.~J. Little (2000).
\newblock Model-based alternatives to trimming survey weights.
\newblock {\em Journal of Official Statistics\/}~{\em 16\/}(3), 191--209.

\bibitem[\protect\citeauthoryear{Firth and Bennett}{Firth and
  Bennett}{1998}]{robust:Firth:JRSSB98}
Firth, D. and K.~E. Bennett (1998).
\newblock Robust models in probability sampling.
\newblock {\em Journal of the Royal Statistical Society Series B\/}~{\em 60},
  3--21.

\bibitem[\protect\citeauthoryear{Fuller}{Fuller}{2009}]{fuller09}
Fuller, W. (2009).
\newblock {\em Sampling Statistics}.
\newblock Hoboken: Wiley.

\bibitem[\protect\citeauthoryear{Gelman}{Gelman}{2005}]{anova:gelman:05}
Gelman, A. (2005).
\newblock Analysis of variance: why it is more important than ever (with
  discusion).
\newblock {\em Annals of Statistics\/}~{\em 33\/}(1), 1--53.

\bibitem[\protect\citeauthoryear{Gelman}{Gelman}{2006}]{gelman06-prior}
Gelman, A. (2006).
\newblock Prior distributions for variance parameters in hierarchical models.
\newblock {\em Bayesian Analysis\/}~{\em 3}, 515--533.

\bibitem[\protect\citeauthoryear{Gelman}{Gelman}{2007}]{gelman07}
Gelman, A. (2007).
\newblock Struggles with survey weighting and regression modeling.
\newblock {\em Statistical Science\/}~{\em 22\/}(2), 153--164.

\bibitem[\protect\citeauthoryear{Gelman and Carlin}{Gelman and
  Carlin}{2001}]{gelmancarlin01}
Gelman, A. and J.~B. Carlin (2001).
\newblock Poststratification and weighting adjustments.
\newblock In R.~Groves, D.~Dillman, J.~Eltinge, and R.~Little (Eds.), {\em
  Survey Nonresponse}.

\bibitem[\protect\citeauthoryear{Gelman and Little}{Gelman and
  Little}{1997}]{gelman:little:97}
Gelman, A. and T.~C. Little (1997).
\newblock Poststratifcation into many cateogiries using hierarchical logistic
  regression.
\newblock {\em Survey Methodology\/}~{\em 23}, 127--135.

\bibitem[\protect\citeauthoryear{Gelman and Little}{Gelman and
  Little}{1998}]{gelman:little:98}
Gelman, A. and T.~C. Little (1998).
\newblock Improving on probability weighting for household size.
\newblock {\em Public Opinion Quarterly\/}~{\em 62}, 398--404.

\bibitem[\protect\citeauthoryear{Ghitza and Gelman}{Ghitza and
  Gelman}{2013}]{Ghitza:gelman-13}
Ghitza, Y. and A.~Gelman (2013).
\newblock Deep interactions with {MRP}: {E}lection turnout and voting patterns
  among small electoral subgroups.
\newblock {\em American Journal of Political Science\/}~{\em 57\/}(3),
  762--776.

\bibitem[\protect\citeauthoryear{Ghosh and Meeden}{Ghosh and
  Meeden}{1997}]{ghosh:meeden:97}
Ghosh, M. and G.~Meeden (1997).
\newblock {\em Bayesian Methods for finite population sampling}.
\newblock Chapman Hall / CRC Press.

\bibitem[\protect\citeauthoryear{Goodrich and Gabry}{Goodrich and
  Gabry}{2017}]{rstanarm}
Goodrich, B. and J.~S. Gabry (2017).
\newblock rstanarm: {B}ayesian applied regression modeling via {S}tan.
\newblock https://cran.r-project.org/web/packages/rstanarm/.

\bibitem[\protect\citeauthoryear{Groves and Couper}{Groves and
  Couper}{1995}]{groves:couper:98JOS}
Groves, R. and M.~Couper (1995).
\newblock Theoretical motivation for post-survey nonresponse adjustment in
  household surveys.
\newblock {\em Journal of Official Statistics\/}~{\em 11}, 93--106.

\bibitem[\protect\citeauthoryear{H{\'a}jek}{H{\'a}jek}{1971}]{hajek71}
H{\'a}jek, J. (1971).
\newblock Comment on ``{A}n essay on the logical foundations of survey
  sampling'' by {D}. {B}asu.
\newblock In V.~P. Godambe and D.~A. Sprott (Eds.), {\em The {F}oundations of
  {S}urvey {S}ampling}, pp.\  236. Holt, Rinehart and Winston.

\bibitem[\protect\citeauthoryear{Hastie, Tibshirani, and Friedman}{Hastie
  et~al.}{2009}]{ml-book09}
Hastie, T., R.~Tibshirani, and J.~Friedman (2009).
\newblock {\em The Elements of Statistical Learning: Data Mining, Inference,
  and Prediction\/} (2nd ed.).
\newblock Springer.

\bibitem[\protect\citeauthoryear{Henry and Valliant}{Henry and
  Valliant}{2012}]{henry:valliant12}
Henry, K. and R.~Valliant (2012).
\newblock Comparing alternative weight adjustment methods.
\newblock In {\em Proceedings of the Section on Survey Research Methods}.
  American Statistical Association.

\bibitem[\protect\citeauthoryear{Hoffman and Gelman}{Hoffman and
  Gelman}{2014}]{hoffman-gelman:2012}
Hoffman, M.~D. and A.~Gelman (2014).
\newblock The {N}o-{U}-{T}urn sampler: {A}daptively setting path lengths in
  {H}amiltonian {M}onte {C}arlo.
\newblock {\em Journal of Machine Learning Research\/}~{\em 15}, 1351--1381.

\bibitem[\protect\citeauthoryear{Holt and Smith}{Holt and Smith}{1979}]{hs79}
Holt, D. and T.~M.~F. Smith (1979).
\newblock Post stratification.
\newblock {\em Journal of the Royal Statistical Society Series A\/}~{\em
  142\/}(1), 33--46.

\bibitem[\protect\citeauthoryear{III and Herriot}{III and
  Herriot}{1979}]{fay:herriot79}
III, R. E.~F. and R.~A. Herriot (1979).
\newblock Estimates of income for small places: An application of james-stein
  procedures to census data.
\newblock {\em Journal of the American Statistical Association\/}~{\em
  74\/}(366a), 269--277.

\bibitem[\protect\citeauthoryear{Kang and Schafer}{Kang and
  Schafer}{2007}]{kang:schafer07}
Kang, J. D.~Y. and J.~L. Schafer (2007).
\newblock Demystifying double robustness: {A} comparison of alternative
  strategies for estimating a population mean from incomplete data.
\newblock {\em Statistical Science\/}~{\em 22\/}(4), 523--539.

\bibitem[\protect\citeauthoryear{Kim and Skinner}{Kim and
  Skinner}{2013}]{Kim:Skinner:BM13}
Kim, J.~K. and C.~J. Skinner (2013).
\newblock Weighting in survey analysis under informative sampling.
\newblock {\em Biometrika\/}~{\em 100}, 385--398.

\bibitem[\protect\citeauthoryear{Kott}{Kott}{2009}]{calibration:kott09}
Kott, P. (2009).
\newblock Calibration weighting: combining probability samples and linear
  prediction models.
\newblock In D.~Pfeffermann and C.~R. Rao (Eds.), {\em Handbook of Statistics,
  Sample Surveys: Design, Methods and Application}, Volume 29B. Elsevier.

\bibitem[\protect\citeauthoryear{Little}{Little}{1983}]{little83-pi}
Little, R. (1983).
\newblock Comment on ``{A}n evaluation of model-dependent and
  probability-sampling inferences in sample surveys", by {M. H. H}ansen, {W. G.
  M}adow and {B. J. T}epping.
\newblock {\em Journal of the American Statistical Association\/}~{\em 78},
  797--799.

\bibitem[\protect\citeauthoryear{Little}{Little}{1991}]{little91}
Little, R. (1991).
\newblock Inference with survey weights.
\newblock {\em Journal of Official Statistics\/}~{\em 7}, 405--424.

\bibitem[\protect\citeauthoryear{Little}{Little}{1993}]{little93}
Little, R. (1993).
\newblock Post-stratification: A modeler's perspective.
\newblock {\em Journal of the American Statistical Association\/}~{\em 88},
  1001--1012.

\bibitem[\protect\citeauthoryear{Little}{Little}{2004}]{little04-model}
Little, R. (2004).
\newblock To model or not to model? {C}ompeting modes of inference for finite
  population sampling inference for finite population sampling.
\newblock {\em Journal of the American Statistical Association\/}~{\em 99},
  546--556.

\bibitem[\protect\citeauthoryear{Little}{Little}{2011}]{CalibratedBayes:Little11}
Little, R. (2011).
\newblock Calibrated {B}ayes, for statistics in general, and missing data in
  particular.
\newblock {\em Statistical Science\/}~{\em 26}, 162--174.

\bibitem[\protect\citeauthoryear{Little and Wu}{Little and
  Wu}{1991}]{rake:little91}
Little, R. and M.~Wu (1991).
\newblock Models for contingency tables with known margins when target and
  sampled populations differ.
\newblock {\em Journal of the American Statistical Association\/}~{\em 86},
  87--95.

\bibitem[\protect\citeauthoryear{McConville and Toth}{McConville and
  Toth}{2018}]{Toth18}
McConville, K.~S. and D.~Toth (2018).
\newblock Automated selection of post-strata using a model-assisted regression
  tree estimator.
\newblock {\em Scandinavian Journal of Statistics\/}~{\em Forthcoming}.

\bibitem[\protect\citeauthoryear{Park, Gelman, and Bafumi}{Park
  et~al.}{2005}]{park:gelman:bafumi-04}
Park, D.~K., A.~Gelman, and J.~Bafumi (2005).
\newblock State-level opinions from national surveys: {P}oststratification
  using multilevel logistic regression.
\newblock In J.~E. Cohen (Ed.), {\em Public Opinion in State Politics}.
  Standord University Press.

\bibitem[\protect\citeauthoryear{Pfeffermann}{Pfeffermann}{1993}]{pfeffermann93}
Pfeffermann, D. (1993).
\newblock The role of sampling weights when modeling survey data.
\newblock {\em International Statistical Review\/}~{\em 61\/}(2), 317--337.

\bibitem[\protect\citeauthoryear{Piironen and Vehtari}{Piironen and
  Vehtari}{2016}]{hyperprior:Aki16}
Piironen, J. and A.~Vehtari (2016).
\newblock On the hyperprior choice for the global shrinkage parameter in the
  horseshoe prior.
\newblock https://arxiv.org/abs/1610.05559.

\bibitem[\protect\citeauthoryear{Potter}{Potter}{1988}]{potter88}
Potter, F.~A. (1988).
\newblock Survey of procedures to control extreme sample weights.
\newblock In {\em Proceedings of the Section on Survey Research Methods}, pp.\
  453--458. American Statistical Association.

\bibitem[\protect\citeauthoryear{Potter}{Potter}{1990}]{trim-potter90}
Potter, F.~A. (1990).
\newblock A study of procedures to identify and trim extreme sampling weights.
\newblock In {\em Proceedings of the Section on Survey Research Methods}, pp.\
  225--230. American Statistical Association.

\bibitem[\protect\citeauthoryear{Rao and Molina}{Rao and Molina}{2015}]{rao15}
Rao, J. and I.~Molina (2015).
\newblock {\em Small Area Estimation}.
\newblock John Wiley \& Sons, Inc.

\bibitem[\protect\citeauthoryear{Rao}{Rao}{1966a}]{Rao66a}
Rao, J. N.~K. (1966a).
\newblock Alternative estimators in {PPS} sampling for multiple
  characteristics.
\newblock {\em Sankhya-Series A\/}~{\em 28\/}(1), 47--60.

\bibitem[\protect\citeauthoryear{Rao}{Rao}{1966b}]{Rao66b}
Rao, J. N.~K. (1966b).
\newblock On the relative efficiency of some estimators in {PPS} sampling for
  multiple characteristics.
\newblock {\em Sankhya-Series A\/}~{\em 28\/}(1), 61--70.

\bibitem[\protect\citeauthoryear{Rao}{Rao}{2011}]{rao:ss11}
Rao, J. N.~K. (2011).
\newblock Impact of frequentist and bayesian methods on survey sampling
  practice: A selective appraisal.
\newblock {\em Statistical Science\/}~{\em 26\/}(2), 240--256.

\bibitem[\protect\citeauthoryear{Rasmussen and Williams}{Rasmussen and
  Williams}{2006}]{gp-rasmussen06}
Rasmussen, C.~E. and C.~K.~I. Williams (2006).
\newblock {\em Gaussian Processes for Machine Learning}.
\newblock MIT Press, Cambridge, MA.

\bibitem[\protect\citeauthoryear{Reilly, Gelman, and Katz}{Reilly
  et~al.}{2001}]{reilly:gelman:katz01}
Reilly, C., A.~Gelman, and J.~Katz (2001).
\newblock Poststratication without population level information on the
  poststratifying variable, with application to political polling.
\newblock {\em Journal of the American Statistical Association\/}~{\em 96},
  1--11.

\bibitem[\protect\citeauthoryear{Royall}{Royall}{1968}]{royall68}
Royall, R.~M. (1968).
\newblock An old approach to finite population sampling theory.
\newblock {\em Journal of the American Statistical Association\/}~{\em 63},
  1269--1279.

\bibitem[\protect\citeauthoryear{Rubin}{Rubin}{1976}]{rubin76}
Rubin, D.~B. (1976).
\newblock Inference and missing data (with discussion).
\newblock {\em Biometrika\/}~{\em 63}, 581--592.

\bibitem[\protect\citeauthoryear{Rubin}{Rubin}{1983}]{rubin83-pi}
Rubin, D.~B. (1983).
\newblock Comment on ``{A}n evaluation of model-dependent and
  probability-sampling inferences in sample surveys," by {M. H. H}ansen, {W. G.
  M}adow and {B. J. T}epping.
\newblock {\em Journal of the American Statistical Association\/}~{\em 78},
  803--805.

\bibitem[\protect\citeauthoryear{S{\"a}rndal, Swensson, and
  Wretman}{S{\"a}rndal et~al.}{1992}]{modelass-sarndal92}
S{\"a}rndal, C.-E., B.~Swensson, and J.~H. Wretman (1992).
\newblock {\em Model Assisted Survey Sampling}.
\newblock Springer, New York.

\bibitem[\protect\citeauthoryear{Si and Gelman}{Si and
  Gelman}{2014}]{RHweighting}
Si, Y. and A.~Gelman (2014).
\newblock Survey weighting for {New York Longitudinal Survey on Poverty
  Measure}.
\newblock Technical report, Columbia University.

\bibitem[\protect\citeauthoryear{Si, Pillai, and Gelman}{Si
  et~al.}{2015}]{bnfp:ba15}
Si, Y., N.~S. Pillai, and A.~Gelman (2015).
\newblock Nonparametric {B}ayesian weighted sampling inference.
\newblock {\em Bayesian Analysis\/}~{\em 10}, 605--625.

\bibitem[\protect\citeauthoryear{Si, Trangucci, and Gabry}{Si
  et~al.}{2017}]{code:si}
Si, Y., R.~Trangucci, and J.~S. Gabry (2017).
\newblock Computation codes for manuscript ``{B}ayesian hierarchical weighting
  adjustment and survey inference''.
\newblock https://github.com/yajuansi-sophie/weighting.

\bibitem[\protect\citeauthoryear{Si and Zhou}{Si and Zhou}{2020}]{BayesRake18}
Si, Y. and P.~Zhou (2020).
\newblock Bayes-raking: Bayesian finite population inference with known
  margins.
\newblock {\em Journal of Survey Statistics and Methodology\/}~{\em
  Forthcoming}.

\bibitem[\protect\citeauthoryear{{Stan Development Team}}{{Stan Development
  Team}}{2017}]{stan-manual:2013}
{Stan Development Team} (2017).
\newblock Stan modeling language user's guide and reference manual.
\newblock http://mc-stan.org.

\bibitem[\protect\citeauthoryear{{Stan Development Team}}{{Stan Development
  Team}}{2018}]{stan-software:2013}
{Stan Development Team} (2018).
\newblock Stan: {A} {C}++ library for probability and sampling.
\newblock http://mc-stan.org.

\bibitem[\protect\citeauthoryear{Tang, Ghosh, Ha, and Sedransk}{Tang
  et~al.}{2018}]{priorSAE-Tang18}
Tang, X., M.~Ghosh, N.~S. Ha, and J.~Sedransk (2018).
\newblock Modeling random effects using global--local shrinkage priors in small
  area estimation.
\newblock {\em Journal of the American Statistical Association\/}~{\em 0\/}(0),
  1--14.

\bibitem[\protect\citeauthoryear{Valliant, Dever, and Kreuter}{Valliant
  et~al.}{2018}]{valliant-book18}
Valliant, R., J.~A. Dever, and F.~Kreuter (2018).
\newblock {\em Practical Tools for Designing and Weighting Survey Samples\/}
  (2nd ed.).
\newblock Springer, New York.

\bibitem[\protect\citeauthoryear{Valliant, Dorfman, and Royall}{Valliant
  et~al.}{2000}]{FPSi:Valliant00}
Valliant, R., A.~Dorfman, and R.~Royall (2000).
\newblock {\em Finite Population Sampling and Inference}.
\newblock Wiley.

\bibitem[\protect\citeauthoryear{Volfovsky and Hoff}{Volfovsky and
  Hoff}{2014}]{volfovsky:hoff14}
Volfovsky, A. and P.~Hoff (2014).
\newblock Hierarchical array priors for {ANOVA} decompositions of
  cross-classified data.
\newblock {\em Annals of Applied Statistics\/}~{\em 8\/}(1), 19--47.

\bibitem[\protect\citeauthoryear{Wang, Rothschild, Goel, and Gelman}{Wang
  et~al.}{2015}]{wang:gelman14}
Wang, W., D.~Rothschild, S.~Goel, and A.~Gelman (2015).
\newblock Forecasting elections with non-representative polls.
\newblock {\em International Journal of Forecasting\/}~{\em 31\/}(3), 980--991.

\bibitem[\protect\citeauthoryear{Wimer, Garfinkel, Gelblum, Lasala, Phillips,
  Si, Teitler, and Waldfogel}{Wimer et~al.}{2014}]{RHreport}
Wimer, C., I.~Garfinkel, M.~Gelblum, N.~Lasala, S.~Phillips, Y.~Si, J.~Teitler,
  and J.~Waldfogel (2014).
\newblock Poverty tracker---monitoring poverty and well-being in {NYC}.
\newblock Columbia Population Research Center and Robin Hood Foundation.

\bibitem[\protect\citeauthoryear{Wu and Sitter}{Wu and
  Sitter}{2001}]{wu:sitter01}
Wu, C. and R.~R. Sitter (2001).
\newblock A model-calibration approach to using complete auxiliary information
  from survey data.
\newblock {\em Journal of the American Statistical Association\/}~{\em
  96\/}(453), 185--193.

\bibitem[\protect\citeauthoryear{Xia and Elliott}{Xia and
  Elliott}{2016}]{elliot:JOS16}
Xia, X. and M.~R. Elliott (2016).
\newblock Weight smoothing for generalized linear models using a {L}aplace
  prior.
\newblock {\em Journal of Official Statistics\/}~{\em 32\/}(2), 507--539.

\bibitem[\protect\citeauthoryear{Yougov}{Yougov}{2017}]{Yougov}
Yougov (2017).
\newblock Introducing the yougov referendum model.
\newblock https://yougov.co.uk.

\bibitem[\protect\citeauthoryear{Yuan and Lin}{Yuan and
  Lin}{2006}]{grouplasso06}
Yuan, M. and Y.~Lin (2006).
\newblock Model selection and estimation in regression with grouped variables.
\newblock {\em Journal of the Royal Statistical Society Series B\/}~{\em 68},
  49--67.

\bibitem[\protect\citeauthoryear{Zhang, Holt, Yun, Lu, Greenlund, and
  Croft}{Zhang et~al.}{2015}]{Zhang15-mrp}
Zhang, X., J.~B. Holt, S.~Yun, H.~Lu, K.~J. Greenlund, and J.~B. Croft (2015).
\newblock Validation of multilevel regression and poststratification
  methodology for small area estimation of health indicators from the
  behavioral risk factor surveillance system.
\newblock {\em American Journal of Epidemiology\/}~{\em 182\/}(2), 127--137.

\end{thebibliography}

\end{document}